\documentclass[useAMS,usenatbib]{mn2e}
\usepackage{epsfig}
\usepackage{amssymb}
\usepackage{amsmath}
\usepackage{times}

\newcommand{\mincir}{\raise
-2.truept\hbox{\rlap{\hbox{$\sim$}}\raise5.truept 
\hbox{$<$}\ }}
\newcommand{\magcir}{\raise
-2.truept\hbox{\rlap{\hbox{$\sim$}}\raise5.truept
\hbox{$>$}\ }}
\newcommand{\minmag}{\raise-2.truept\hbox{\rlap{\hbox{$<$}}\raise
6.truept\hbox
{$>$}\ }}

\newcommand{\lya}{Lyman-$\alpha$~}
\newcommand{\gad} {{\small {GADGET-2}}\,}
\newcommand{\lyb}{Lyman-$\beta$~}

\newcommand{\be}{\begin{equation}}
\newcommand{\ee}{\end{equation}}
\newcommand{\ba}{\begin{eqnarray}}
\newcommand{\ea}{\end{eqnarray}}
\newcommand{\brr}{\begin{array}}
 
\newcommand{\err}{\end{array}}
\newcommand{\bc}{\begin{center}}
\newcommand{\ec}{\end{center}}

\DeclareMathAlphabet{\mathsc}{OT1}{cmr}{m}{sc}
\def\testbx{bx}%
\DeclareRobustCommand{\ion}[2]{%
\relax\ifmmode
\ifx\testbx\f@series
{\mathbf{#1\,\mathsc{#2}}}\else
{\mathrm{#1\,\mathsc{#2}}}\fi
\else\textup{#1\,{\mdseries\textsc{#2}}}%
\fi}

\title[Cosmic evolution of the CIV]{Cosmic evolution of the CIV in
  high-resolution hydrodynamic simulations}

 \author[E. Tescari et al.]{E. Tescari$^{1,2,3}$\thanks{E-mail:
     tescari@oats.inaf.it}, M. Viel$^{1,2}$, V. D'Odorico$^{1}$,
   S. Cristiani$^{1,2}$, F. Calura$^{1,4}$, S. Borgani$^{1,2,3}$
   \newauthor{and L. Tornatore$^{1,2,3}$} \\ $^1$ INAF - Osservatorio
   Astronomico di Trieste, Via G.B. Tiepolo 11, I-34131 Trieste, Italy
   \\ $^2$ INFN/National Institute for Nuclear Physics, Via Valerio 2,
   I-34127 Trieste, Italy \\ $^3$ Dipartimento di Fisica - Sezione di
   Astronomia, Universit\`a di Trieste, Via G.B. Tiepolo 11, I-34131
   Trieste, Italy \\ $^4$ Jeremiah Horrocks Institute for Astrophysics
   and Supercomputing, University of Central Lancashire, Preston PR1
   2HE}

\begin{document}

\maketitle

\begin{abstract}

We investigate the properties of triply ionized Carbon (CIV) in the
Intergalactic Medium (IGM) using a set of high-resolution and large
box-size cosmological hydrodynamic simulations of a $\Lambda$CDM
model. We rely on a modification of the public available Tree-Smoothed
Particle Hydrodynamics (SPH) code {\small{GADGET-2}} that
self-consistently follows the metal enrichment mechanism by means of a
detailed chemical evolution model. We focus on several numerical
implementations of galactic feedback: galactic winds in the energy
driven and momentum driven prescriptions, galactic winds
hydrodynamically coupled to the surrounding gas and Active Galactic
Nuclei (AGN) powered by gas accretion onto massive black
holes. Furthermore, our results are compared to a run in which
galactic feedback is not present and we also explore different initial
stellar mass function. After having addressed some global properties
of the simulated volume like the impact on the star formation rate and
the content in Carbon and CIV, we extract mock IGM transmission
spectra in neutral hydrogen (HI) and CIV and perform Voigt profile
fitting. The results are then compared with high-resolution quasar
(QSO) spectra obtained with the UVES spectrograph (Ultra Violet
Echelle Spectrograph) at the VLT (Very Large Telescope) and the High Resolution Echelle Spectrometer (HIRES) at Keck. We find that feedback has little impact on
statistics related to the neutral hydrogen, while CIV is more affected
by galactic winds and/or AGN feedback. The feedback schemes
investigated have a different strength and redshift evolution with a
general tendency for AGN feedback to be more effective at lower
redshift than galactic winds. When the same analysis is performed over
observed and simulated CIV lines, we find reasonables good agreement
between data and simulations over the column density range $N_{\rm
  CIV}=10^{12.5-15}$ cm$^{-2}$. Also the CIV line-widths distribution
appears to be in agreement with the observed values, while the HI
Doppler parameters, $b_{\rm HI}$, are in general too large showing
that the diffuse cosmic web is heated more than what is inferred by
observations. The simulation without feedback fails in reproducing the
CIV systems at high column densities at all redshift, while the AGN
feedback case agrees with observations only at $z<3$, when this form
of feedback is particularly effective. We also present scatter plots
in the $b-N$ and in the $N_{\rm CIV}-N_{\rm HI}$ planes, showing that
there is rough agreement between observations and simulations only
when feedback is taken into account.

Although it seems difficult to constrain the nature and the strength
of galactic feedback using the present framework and find a unique
model that fits all the observations, these simulations offer the
perspective of understanding the galaxy-IGM interplay and how metals
produced by stars can reach the low density IGM.

\end{abstract}


\section{Introduction}

The interplay between galaxies and the intergalactic medium (IGM) is a
fundamental and poorly understood aspect in the standard scenario of
structure formation. The intimate connection between the diffuse IGM
and the baryons that populate star forming galaxies can shed light on
the metal enrichment of the IGM (its nature and redshift evolution),
on the properties of galactic feedback and on the evolution of the
ultraviolet background.

In recent years, this field has benefitted from spectroscopical
observations of quasar (QSO) and Gamma Ray Burst (GRB) spectra (both
at high and low resolution) and from photometric and spectroscopical
information on high redshift galaxies in the same field of view. This
has in turn allowed us to perform cross-correlation studies of absorption
lines and properties of galaxies \citep[e.g.][and references
  therein]{steidel010}. Observations have revealed that the IGM is
polluted with heavy elements such as Carbon, Oxygen and Silicon
\citep[e.g.][]{weymann81,cowie95, aguirre05}. Furthermore, high
redshift star forming galaxies have been shown to interact with the
surrounding medium by powering strong galactic outflows
\citep[e.g.][]{adelberger05}. The IGM enrichment mechanism is,
however, still poorly constrained and the extent of metal pollution of
the low density IGM, or around Lyman-break galaxies, controversial
\citep[e.g.][]{schaye00,pieri06,porciani05}.

Among many different metal species, the doublet produced by triply
ionized carbon (CIV) is the one that has been most widely studied
since it is relatively easy to identify in the regions redward of \lya
emission. CIV has been observed from low-redshift up to very high
redshift \citep{ryan-weber06,ryan-weber09,becker2009}; its clustering
properties in the transverse direction and along the line-of-sight
have been investigated \citep{vale02,pichon03,tytler09} along with the
global evolution of the CIV density fraction down to $z<2$
\citep{vale10}; the optical depth in CIV has been derived using
pixel-optical-depth techniques applied to the UVES Large Programme
sample in order to constrain the metallicity of the low density gas
\citep{schaye03,aracil04}; the cross-correlation between galaxies and
CIV properties have also been addressed in great detail using either
Keck or UVES spectra \citep{simcoe06,scanna06}. The general conclusion
is that CIV absorption systems are indeed related to outskirts of
galactic haloes but it is still unclear whether the gas producing the
absorption is enriched by nearby galaxies or at an earlier epoch by
smaller objects.

The large amount of data available has prompted the implementation of
relevant physical processes into sophisticated hydrodynamic codes in
order to simulate both galaxies and QSO/GRB absorption lines. The
first attempt to model CIV properties using hydrodynamic simulations
has been made by \citet{haehnelt96}. Subsequently, it has been
demonstrated that a key ingredient that appears to be necessary in
order to reproduce CIV metal line statistics is feedback in the form
of galactic winds \citep{theuns02,cen05,oppe06,oppe07}. Galactic winds
triggered by supernovae (SN) explosions have been modelled in a simple
way using mainly two schemes that conserve either energy or momentum
\citep{oppe06,tex09}. Other sources of feedback like Active Galactic
Nuclei (AGN) has also been considered to address global gas properties
from high to low redshift \citep{tornatore2010}. A second very
important ingredient has been the implementation of accurate chemical
evolution models that self-consistently follow the release of
different metal species from stars and also allow for different
initial stellar mass functions
\citep{tornatore07,kawata07,koba07,sommer08,wiersma09,oppe09,wiersma10,dave010,cen010}.

This is a second paper in a series that exploits the capabilities of a
state-of-the-art modification of the public available code {\small
  GADGET-2} \citep{springel2005}. In the first paper we addressed
neutral hydrogen properties \citep{tex09} focusing in particular on
Damped Lyman-$\alpha$ systems. In this paper we study the CIV
transition by extending the feedback models simulated.

This paper is organized as follows. In Sections \ref{obdatasamp} and
\ref{sec_sim} we present, respectively, the observational data sample
and our set of simulations along with the different feedback models
implemented in our code. In Section \ref{sec:sfrprop} we compare some
global properties of the simulations: the cosmic star formation rate
and the evolution of the total Carbon content. We then focus on the
neutral hydrogen statistics in Section \ref{sec_histati}. In Sections
\ref{sec_civcddf} and \ref{sec:omcivev} we investigate the CIV column
density distribution function and the evolution with redshift of the
CIV cosmological mass density, $\Omega_{\rm CIV}$, respectively. In
Section \ref{sec:bcivpdf} we study the probability distribution
function of the CIV Doppler parameter and, in Section
\ref{sec:bcivNciv}, we present the CIV column density-Doppler
parameter relation. The last part of the paper is dedicated to the
analysis of the HI-CIV correlated absorption (Section
\ref{sec_corrCIVHI}). Finally, in Section \ref{concl} we summarize our
main results and we draw some conclusions.

\section{Observational data sample}
\label{obdatasamp}

The core of the observational data sample is formed by the high
resolution ($R\sim 45000$), high signal-to-noise (S/N) QSO spectra
obtained with the UVES spectrograph \citep{dekker2000} at the Kueyen
unit of the ESO VLT (Cerro Paranal, Chile) in the framework of the ESO
Large Programme (LP): ``The Cosmic Evolution of the IGM''
\citep{bergeron04}.

We have used the sample of HI \lya absorption lines described in
\citet{saitta08} and \citet{vale08}, which is based on 18 objects of
the LP plus 4 more QSOs of which: J2233-606 and HE1122-1648 were
observed with UVES, while HS1946+7658 and B1422+231 were obtained with the High Resolution Echelle Spectrometer (HIRES) at the Keck Telescope at comparable resolution
and signal-to-noise ratio. In this work, also the HI \lya list of the
QSO Q0055-269, observed with UVES, has been added to the sample. The
adopted sample of CIV absorptions was described in \citet{vale10}. The
previous group of QSOs was increased with the addition of the UVES
spectra of PKS2000-330 and PKS1937-101 to extend the redshift interval
to $z\;\magcir 3$.
 
All UVES spectra were reduced with the UVES pipeline following the
standard procedure. The continuum level was determined by
interpolating with a cubic spline the region of the spectrum free from
evident absorption features. Absorption lines were fitted with Voigt
profiles using the context {\small LYMAN} of the {\small MIDAS}
reduction package \citep{font&ball95}. A minimum number of components
was adopted to fit the velocity profile in order to reach a normalized
$\chi^2\sim 1$. For all the QSO in the sample, the \lya forest was
defined as the interval between the \lyb emission plus 1000 km
s$^{-1}$ (to avoid contamination from the \lyb forest), and 5000 km
s$^{-1}$ from the \lya emission; the CIV forest was defined as the
interval between the \lya emission and 5000 km s$^{-1}$ from the CIV
emission. The higher bound was fixed to avoid the proximity region
affected by the QSO emission, where most of the intrinsic systems are
found.

\begin{table}
\begin{center}
\begin{tabular}{l l c c c}
\hline
QSO & $z_{\rm em}$ & $\Delta z_{\rm Ly\alpha}$ & $\Delta z_{\rm C IV}$  \\
\hline
HE 1341-1020 & 2.142  & 1.6599-2.090  & 1.467-2.090 \\
Q0122-380   & 2.2004  & 1.709-2.147  & 1.513-2.147 \\
PKS 1448-232 & 2.224  & 1.729-2.171  & 1.531-2.171 \\
PKS 0237-23  & 2.233  & 1.737-2.179  & 1.538-2.179 \\
J2233-606   & 2.248   & 1.7496-2.194 & 1.550-2.194 \\
HE 0001-2340 & 2.265  & 1.764-2.211  & 1.564-2.211 \\
HS 1626+6433$^a$ & 2.32  & & 1.607-2.265 \\
HE 1122-1648$^b$ & 2.40  & 1.878-2.344  & 1.665-2.344 \\
Q0109-3518   & 2.4057   & 1.883-2.349  & 1.674-2.349 \\
HE 2217-2818  & 2.414   & 1.888-2.355   & 1.681-2.355 \\
Q0329-385    & 2.435    & 1.9096-2.378  & 1.697-2.378 \\
HE 1158-1843  & 2.448   & 1.919-2.391   & 1.707-2.391\\
HE 1347-2457  & 2.5986  & 2.046-2.539  & 1.826-2.539 \\
Q1442+2931$^a$   & 2.661 & & 1.875-2.600 \\
Q0453-423    & 2.669  & 2.106-2.608  & 1.881-2.608 \\
PKS 0329-255  & 2.696  & 2.129-2.635 & 1.902-2.635 \\
HE 0151-4326  & 2.763  & 2.186-2.701 & 1.955-2.701 \\
Q0002-422    & 2.769   & 2.191-2.707 & 1.959-2.707 \\
HE 2347-4342  & 2.880  & 2.285-2.816 & 2.067-2.816 \\
SBS 1107+487$^a$  & 2.966 & & 2.114-2.900 \\
HS 1946+7658$^c$  & 3.058  & 2.435-2.991 & 2.181-2.991 \\
HE 0940-1050  & 3.0932 & 2.465-3.025  & 2.214-3.025 \\
Q0420-388    & 3.1257  & 2.493-3.057 & 2.239-3.057 \\
S4 0636+68$^a$ & 3.175 & & 2.278-3.106 \\
SBS 1425+606$^a$ & 3.199 & & 2.297-3.129 \\
PKS 2126-158 & 3.292   & 2.633-3.221 & 2.370-3.221 \\
B1422+231$^b$   & 3.623 & 2.914-3.546 & 2.630-3.546 \\
Q0055-269   & 3.66  & 2.945-3.583 & 2.659-3.583\\
PKS 2000-330 & 3.783   && 2.756-3.704 \\
PKS 1937-101 & 3.787   && 2.770-3.400 \\
PSS J1646+5514$^a$ & 4.059 && 2.972-3.975 \\
PSS J1057+4555$^a$ & 4.131 && 3.029-4.046 \\
BR 2237-0607$^a$   & 4.559 && 3.365-4.467 \\
\hline
\end{tabular}
\caption{Relevant properties of the QSOs forming the \citet{vale10}
  observational sample. $(a)$ QSOs from \citet{boksenberg2003}; $(b)$
  Reduced spectra kindly provided by T.-S. Kim; $(c)$ Data from
  \citet{kirk97}.}
\end{center}
\label{tab_obs}
\end{table}

CIV features are distinguished into {\it components} or {\it lines},
meaning the velocity components in which every absorption profile has
been decomposed, and {\it systems}, formed by groups of components
with relative separation smaller than $dv_{\rm min} = 50$ km s$^{-1}$
\citep[see][for a detailed description of the procedure]{vale10}. Both
the velocity components and the systems in the sample have column
densities in the range $10^{11}\,{}^<_{\rm \sim}\, N_{\rm
  CIV}\,{}^<_{\rm \sim}\, 10^{15}$ cm$^{-2}$.

In order to further increase the number of CIV lines and to extend the
sample to higher redshift, \citet{vale10} have considered the CIV
lines fitted in 9 QSO spectra observed with HIRES at Keck at a
resolution and signal-to-noise ratio similar to those of their spectra
and reported in \citet{boksenberg2003}. The fit with Voigt profiles
was carried out by \citet{boksenberg2003} with {\small
  VPFIT}\footnote{http://www.ast.cam.ac.uk/$\sim$rfc/vpfit.html}. The
main difference between {\small LYMAN} and {\small VPFIT} is that the
number of components fitted to a given velocity profile is, in
general, larger using the latter \citep[see also the discussion
  in][]{saitta08}. This is seen also in the analysis of
\citet{vale10}, in particular from the comparison of the CIV lines
detected in the spectrum of the QSO B1422+231, which is the only
object in common between the two samples. \citet{vale10} find that in
all cases the number of components determined with {\small VPFIT} is
larger or equal to that determined with {\small LYMAN}. However, when
the total column density of each absorption system is considered the
difference between the two fitting procedures becomes negligible.

All the QSOs of the observational sample are reported in Table 1 with
their emission redshift and the redshift range covered by the \lya
forest and the CIV lines.

\section{The simulations}
\label{sec_sim}

In the following we review the main characteristics of the runs
analysed in this work and we refer to \citet{tex09} for a more
extensive description of the simulations. Our code is the same
modified version of the TreePM-SPH code \gad \citep{springel2005} used
in \citet{tex09}.

\begin{table*}
\centering
\begin{tabular}{llcccccc}
\\ \hline & Run & IMF & Box Size & \textit{N$_{\rm GAS}$}& \textit{m$_{\rm GAS}$} &
Softening & Feedback \\ & & & [$h^{-1}$Mpc] &
& [$h^{-1}M_{\rm \odot}$] & [$h^{-1}$kpc] \\ \hline
& kr37edw500 & Kroupa & 37.5 & $256^3$ & $3.6 \times 10^7$ &  7.5 & EDW \\
& kr37mdw & Kroupa & 37.5 & $256^3$ & $3.6 \times 10^7$ &  7.5 & MDW$^a$ \\
& kr37agn & Kroupa & 37.5 & $256^3$ & $3.6 \times 10^7$ &  7.5 & AGN$^b$ \\
& kr37agn+edw300 & Kroupa & 37.5 & $256^3$ & $3.6 \times 10^7$ &  7.5 & AGN+EDW$^c$ \\
& kr37nf & Kroupa & 37.5 & $256^3$ & $3.6 \times 10^7$ &  7.5 & NO FEEDBACK \\
& kr37co-edw500 & Kroupa & 37.5 & $256^3$ & $3.6 \times 10^7$ &  7.5 & COUPLED EDW$^d$ \\
& ay37edw500 & Arimoto-Yoshii & 37.5 & $256^3$ & $3.6 \times 10^7$ &  7.5 & EDW \\
& sa37edw500 & Salpeter & 37.5 & $256^3$ & $3.6 \times 10^7$ &  7.5 & EDW \\
& kr37p400edw500 & Kroupa & 37.5 & $400^3$ & $9.4 \times 10^6$ &  4.8 & EDW \\
\hline \\
\end{tabular}
\caption{Summary of the different runs. Column 1, run name; column 2,
  Initial Mass Function (IMF) chosen; column 3, comoving box size;
  column 4, number of gas particles; column 5, mass of gas particle;
  column 6, Plummer-equivalent comoving gravitational softening;
  column 7, type of feedback implemented. For the runs kr37edw500,
  kr37co-edw500, ay37edw500, sa37edw500 and kr37p400edw500 the
  velocity of the energy-driven winds (EDW) is set to $v_{\rm w}=500$
  km s$^{-1}$.  $(a)$: momentum-driven winds (MDW) feedback in which
  wind velocity scales roughly with $3\sigma$ ($\sigma$ being the
  velocity dispersion of the halo that hosts the ``wind''
  particle). $(b)$: AGN feedback in which the energy is released by
  gas accretion onto super-massive black holes. $(c)$: combined effect
  of energy-driven winds of 300 km s$^{-1}$ and AGN feedbacks. $(d)$:
  energy-driven winds not decoupled from the hydrodynamics (see
  Section \ref{subsec_endr}).}
\label{tab:sim_civ}
\end{table*}

The simulations cover a cosmological volume (with periodic boundary
conditions) filled with an equal number of dark matter and gas
particles. The cosmological model chosen is a flat $\Lambda$CDM with
the following parameters: $\Omega_{\rm 0m}=0.24$, $\Omega_{\rm
  0b}=0.0413$, $\Omega_{\rm \Lambda}=0.76$, $n_{\rm s}=0.96$, $H_{\rm
  0}=73$ km s$^{-1}$ Mpc$^{-1}$ and $\sigma_{\rm 8}=0.8$, which are in
agreement with the latest results from large scale structure
observables such as the cosmic microwave background, weak lensing, the
\lya forest and the evolution of mass function of galaxy clusters
\citep{lesg08,komatsu08,vikhetal09}.  The input linear dark matter
power spectrum for the initial conditions has been generated at $z=99$
with {\small CMBFAST} \citep{seljakzalda96} and includes baryonic
acoustic oscillations.

Radiative cooling and heating processes are modelled following the
implementation of \citet{KWH}. We assume a mean Ultra Violet
Background (UVB) produced by quasars and galaxies as given by
\citet{HM96}, with the heating rates multiplied by a factor of $3.3$ in
order to better fit observational constraints on the temperature
evolution of the intergalactic medium at high redshift \citep[the
  factor of $3.3$ was originally introduced because it gives a
  temperature at the mean cosmic density, $T_{\rm 0}$, in agreement
  with the][data points]{schaye00b}. This background gives naturally
$\Gamma_{\rm -12} \sim 0.8-1$ at low redshift, in agreement with
recent observational measurements \citep{bolton05,fg08}. Multiplying
the heating rates by the factor above (chosen empirically) results in
a larger IGM temperature at the mean density which cannot be reached
by the standard hydrodynamic code. This aims at mimicking, at least in
a phenomenological way, the non-equilibrium ionization effects around
reionization \citep[see for example][]{ricotti00,schaye00,Bol07}. A
significant fraction in mass of the IGM resides in the tight power-law
relation $T=T_{\rm 0}(1+\delta)^{\gamma-1}$ at around the mean
density, where $T_{\rm 0}$ is the temperature at the mean cosmic
density and $\delta$ the overdensity. Our reference simulations (see
at the end of this Section), at redshift $z=3$, 2.25 and 1.5, have
$\log \langle T_{\rm 0}$(K)$\rangle$, respectively, equal to: 4.36,
4.33, 4.30 and $\gamma$, respectively, equal to: 1.58, 1.60, 1.62.

In our simulations, the standard multiphase star formation criterion
of \citet{springel2003} is implemented, which contains an effective
prescription for the inter-stellar medium (ISM). In this effective
model, whenever the density of a gas particle exceeds a given
threshold (set to $\rho_{\rm th}=0.1$ cm$^{-3}$ in terms of the number
density of hydrogen atoms), that gas particle is flagged as star
forming and is treated as multiphase. With this prescription baryons
are in the form either a hot or a cold phase or in stars, thereby this
density threshold marks the onset of cold clouds formation.

We follow self-consistently the evolution of these elements: H, He, C,
O, Mg, S, Si and Fe. The contribution of metals is included in the
cooling function using the tables of \citet{SD93}, that assume the
solar value for the relative abundances. In this paper we use the
solar metallicity and element abundances given by \citet{asplund05},
with $Z/X=0.0165$. Besides including different contributions from Type
II and Type Ia supernovae (SN II, SN Ia) and Low and Intermediate-Mass
Stars (LIMS), our model of chemical evolution accounts for
mass-dependent stellar lifetimes. We adopt the lifetime function given
by \citet{padovanimatteucci93}. For the stellar yields we use: SNIa --
\citet{thielemann03}; SNII -- \citet{woosleyweaver95}; LIMS --
\citet{vandenhoek97}. The mass-range for the SNII is $m > 8 M_{\rm
  \odot}$, while for the SNIa is $m < 8 M_{\rm \odot}$ with a binary
fraction of 10\%. Finally we use three distinct stellar initial mass
functions (IMFs): a Kroupa, a Salpeter and an Arimoto-Yoshii IMF. For
this paper our reference choice is the Kroupa IMF \citep{Kroupa01},
which consists of a multi-slope approximation: $y_{\rm KR}=0.3$ for
stellar mass $m < 0.5M_{\rm \odot}$, $y_{\rm KR}=1.2$ for $0.5M_{\rm
  \odot} \leq m < 1M_{\rm \odot}$ and $y_{\rm KR}=1.7$ for $m \geq
1M_{\rm \odot}$ (where $\varphi(m) \propto m^{\rm -y}$ is the
functional form for the IMF and $\varphi(m)$ is the IMF by
mass). Salpeter \citep{salpeter55} IMF has single slope $y_{\rm
  SL}=1.35$ and Arimoto-Yoshii \citep{ay87} IMF also has single slope
$y_{\rm AY}=0.95$.

In this paper we use three different feedback schemes: two associated
to the galactic (energy and momentum driven) winds produced by
``starburst'' galaxies, already presented in \citet{tex09}, and the
(new) AGN feedback associated to the energy released by gas accretion
onto super-massive black holes. In the following we briefly describe
these different models.

\subsection{Energy-driven winds}
\label{subsec_endr}

For this implementation, the wind mass-loss rate $\dot{M}_{\rm w}$ is
assumed to be proportional to the star formation rate $\dot{M}_{\rm
  \star}$ according to
\begin{equation}
  \dot{M}_{\rm w}=\eta\dot{M}_{\rm \star}, 
\end{equation}
where $\eta$ is the wind mass loading factor (i.e. the wind
efficiency), and the wind carries a fixed fraction $\chi$ of SN
energy:
\begin{equation}
  \frac{1}{2}\dot{M}_{\rm w}v_{\rm w}^2=\chi\epsilon_{\rm SN}\dot{M}_{\rm \star}.
\end{equation}
Star forming gas particles are then stochastically selected according
to their star formation rate to become part of a blowing
wind. Whenever a particle is uploaded to the wind, it is decoupled
from the hydrodynamics for a given period of time or till the density
around it drops below a given fraction of the density threshold for
the onset of the star formation ($\rho_{\rm th}=0.1$ cm$^{-3}$ in
terms of the number density of hydrogen atoms), in order to
effectively reach less dense regions. The (lower) limiting density for
decoupling is set to $0.5\rho_{\rm th}$. This allows the wind particle
to travel ``freely'' up to few kpc until it has left the dense star
forming phase, without directly interacting with it. Otherwise, to
verify the effect of the decoupling, we run a ``coupled winds''
simulation in which wind particles are always affected by the
hydrodynamics like all the other gas particles \citep[see
  also][]{dvecchia&schaye08,tornatore2010}. Unlike in
\citet{springel2003}, we fixed the velocity of the winds, $v_{\rm w}$,
instead of the fraction of the energy made available by SNII
explosions to power galactic ejecta. Thus, four parameters fully
specify the wind model: the wind efficiency $\eta$, the wind speed
$v_{\rm w}$, the wind free travel length $l_{\rm w}$ and the wind free
travel density factor $\delta_{\rm w}$. The maximum allowed time for a
wind particle to stay hydrodynamically decoupled is then $t_{\rm dec}
= l_{\rm w} / v_{\rm w}$ (note that in the case of coupled winds
$t_{\rm dec}=0$). The parameter $l_{\rm w}$ has been introduced in
order to prevent a gas particle from being trapped into the potential
well of the virialized halo and in order to effectively escape from
the ISM, reach the low density IGM and pollute it with metals.

In this paper, we use $l_{\rm w}=10$ $h^{-1}$ kpc and we considered
two different values for the wind velocity: $v_{\rm w}=300$ and 500 km
s$^{-1}$. In our implementation the parameter $\eta$ is kept fixed to
the value 2.

\subsection{Momentum-driven winds}

A feasible physical scenario for the wind driving mechanism, outlined
by \citet{Murrayetal2005} to explain observational evidence of
outflows from starburst galaxies \citep{Martin2005,Rupkeetal2005}, is
the so called momentum-driven wind model. In this scenario, the
radiation pressure of the starburst drives an outflow, possibly by
transferring momentum to an absorptive component (e.g. dust) that is
hydrodynamically coupled with the gas component, which is then dragged
out of the galaxy. Following \citet{oppe06,oppe07} we implemented the
momentum-driven winds model in our code. In such a model the wind
speed scales as the galaxy velocity dispersion $\sigma$, as observed
by \citet{Martin2005}. Since in momentum-driven winds the amount of
input momentum per unit star formation is constant, this implies that
the wind mass loading factor $\eta$ must be inversely proportional to
the velocity dispersion. We therefore use the following relations:
\begin{equation}
\label{eqwind}
v_{\rm wind}=3\sigma\sqrt{f_{\rm L}-1};\ \ \ \ \eta=\frac{\sigma_{\rm 0}}{\sigma},
\end{equation}
where $f_{\rm L}$ is the luminosity factor in units of the galactic
Eddington luminosity (i.e. the critical luminosity necessary to expel
gas from the galaxy), $\sigma_{\rm 0}$ is the normalization of the
mass loading factor and we add an extra $2\sigma$ kick to get out of
the potential of the galaxy in order to simulate continuous pumping of
gas until it is well into the galactic halo. To determine $\sigma$, we
identify haloes runtime in the simulations by using a parallel
friends-of-friends (FoF) algorithm and we associate $\sigma$ with the
velocity dispersion of the haloes, $\sigma_{\rm DM}$ \citep[for
  further details see Section 2.2 of][]{tex09}. \citet{schaerer03}
suggested an approximate functional form for far-UV emission which
\citet{oppe06,oppe07} used to obtain the luminosity factor and which
includes a metallicity dependence for $f_{\rm L}$, owing to more UV
photons output by lower-metallicity stellar populations:
\begin{equation}
f_{\rm L}=f_{\rm L,\odot}\times10^{-0.0029(\log Z+9)^{2.5}+0.417694},
\end{equation}
where \citet{Martin2005} suggests: $f_{\rm L,\odot}\approx2$. The mass
loading factor controls star formation at early times, so $\sigma_{\rm
  0}$ can also be set by requiring a match to the observed global star
formation rate. Following \citet{oppe06,oppe07} we set $\sigma_{\rm
  0}=150$ km s$^{-1}$. Even for this wind implementation the particles
are stochastically selected in the same way as for the energy-driven
scenario.

\subsection{AGN feedback}
\label{agn_feedb}

Finally, we include in our simulations the effect of feedback energy
from gas accretion onto super-massive black holes (BHs), following the
scheme originally introduced by \citet{springetal05} \citep[see
  also][]{dimatteo05}. In this model, BHs are represented by
collisionless sink particles initially seeded in dark matter haloes,
which subsequently grow via gas accretion and through mergers with
other BHs during close encounters. Every new dark matter halo,
identified by a run-time friends-of-friends algorithm, above the mass
threshold $M_{\rm th}=10^{10}$ $M_{\rm \odot}$, is seeded with a
central BH of initial mass $10^5$ $M_{\rm \odot}$, provided the halo
does not contain any BH yet. Each BH can then grow by local gas
accretion, with a rate given by
\begin{equation}
\dot M_{\rm BH}={\rm min}\left(\dot M_{\rm B}, \dot M_{\rm
  Edd}\right),
\label{eq:acrate}
\end{equation} 
where $\dot M_{\rm B}$ is the accretion rate estimated with the
Bondi-Hoyle-Littleton formula \citep[e.g.][]{bondi1952} and $\dot
M_{\rm Edd}$ is the Eddington accretion rate. We refer to
\citet{springetal05} for the details of the AGN feedback
implementation and to \citet{booth&schaye09} for a review of the black
holes accretion feedback in the context of cosmological
simulations. An important parameter of the model is the radiative
efficiency, $\epsilon_{\rm r}$, which gives the radiated energy
$L_{\rm r}$ in units of the energy associated to the accreted mass:
\begin{equation}
\epsilon_{\rm r}=\frac{L_{\rm r}}{\dot M_{\rm BH} c^2}.
\end{equation}
Following \citet{springetal05} and \citet{dunja2010}, we fix
$\epsilon_{\rm r}=0.1$ as a reference value, which is typical for a
radiatively efficient accretion onto a Schwarzschild BH
\citep{shakura1973}. The model then assumes that a fraction
$\epsilon_{\rm f}$ of the radiated energy is thermally coupled to the
surrounding gas, so that $\dot E_{\rm feed}=\epsilon_{\rm r}
\epsilon_{\rm f} \dot M_{\rm BH}c^2$ is the rate of the energy
released to heat the surrounding gas. Using $\epsilon_{\rm f}\sim
0.05$, \citet{dimatteo05} were able to reproduce the observed $M_{\rm
  BH}-\sigma$ relation between bulge velocity dispersion and mass of
the hosted BH \citep[see also][]{sizicki2008,dimatteo08}. Gas particle
accretion onto the BH is implemented in a stochastic way, by assigning
to each neighbouring gas particle a probability of contributing to the
accretion, which is proportional to the SPH kernel weight computed at
the particle position. In the scheme described above, this stochastic
accretion is used only to increase the dynamic mass of the BHs, while
the mass entering in the computation of the accretion rate is followed
in a continuous way, by integrating the analytic expression for $\dot
M_{\rm BH}$. Once the amount of energy to be thermalized is computed
for each BH at a given timestep, this energy is then distributed to
the surrounding gas particles using the SPH kernel weighting scheme.
\begin{figure*}
\centering
\includegraphics[width=16cm,height=8cm]{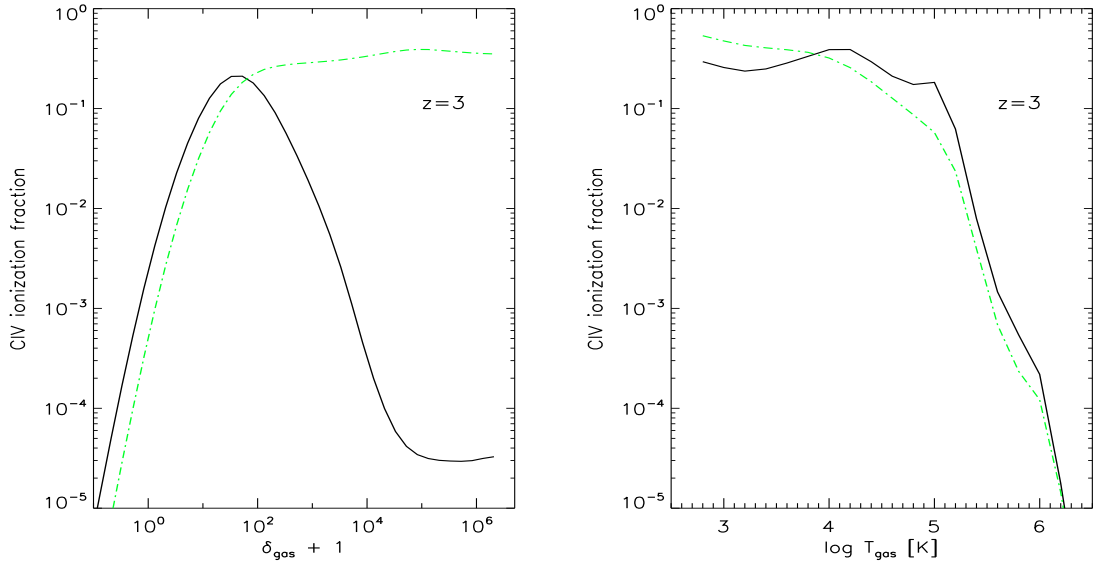}
\caption{\textit{Left panel}: CIV ionization fraction, as a function
  of the gas overdensity, at $z=3$ for a gas temperature of 10$^{4.6}$
  K (solid black line) and 10$^{5.0}$ K (dot-dashed green
  line). \textit{Right panel}: CIV ionization fraction, as a function
  of the gas temperature, at $z=3$ for an overdensity of $\delta_{\rm
    gas}\sim 50$ (solid black line) and $\delta_{\rm gas}\sim 10$
  (dot-dashed green line).}
\label{fig:CIV_hm05new}
\end{figure*}
Our research group explored for the first time this new feedback
prescription in relation to the high redshift properties of the
intergalactic medium, while \citet{tornatore2010} have studied the
impact of this feedback mechanism on the low redshift IGM \citep[other
  groups are working on the black holes accretion feedback related to
  IGM and the Warm Hot Intergalactic Medium (WHIM), see for
  example][]{wiersma10,bertone10arXiv,bertone10mnras}.\\

In Table \ref{tab:sim_civ} we summarize the main parameters of the
cosmological simulations performed, including the mass associated to
the gas particles and the gravitational softening. All the simulations
start at redshift $z=99$ and stop at redshift $z=1.5$. In the
following we outline the specific characteristics of each run:
\begin{itemize}
\item {\bf kr37edw500}: Energy-driven winds (EDW) of velocity $v_{\rm
  w}=500$ km s$^{-1}$ and Kroupa IMF (the reference IMF for this
  work).
\item {\bf kr37mdw}: Momentum-driven winds (MDW) and Kroupa IMF.
\item {\bf kr37agn}: In this simulation the AGN feedback mechanism,
  associated to the energy released by gas accretion onto
  super-massive black holes and described in Section \ref{agn_feedb},
  is active.
\item {\bf kr37agn+edw300}: In this simulation two different feedback
  models are combined: the energy-driven galactic winds of velocity
  $v_{\rm w}=300$ km s$^{-1}$, and the AGN feedback. We will see that
  this run is completely different from the kr37agn simulation
  mentioned above. In fact, the galactic winds start to be effective
  at higher redshift than the AGN feedback, making the haloes devoid of gas and consequently reducing the efficiency of the black holes
  accretion and the power of the AGN feedback. Therefore, the effect
  of the winds is more prominent than that of the AGN, with the result
  that the kr37agn+edw300 run behaves very similarly to a regular
  energy-driven winds simulation.
\item {\bf kr37nf}: This simulation was run without any winds or AGN
  feedback in order to test how large the effects of the
  different feedback prescriptions on the CIV statistics are.
\item {\bf kr37co-edw500}: as we stated in Section \ref{subsec_endr},
  this simulation has energy-driven winds of velocity $v_{\rm w}=500$
  km s$^{-1}$, never decoupled from the hydrodynamics. We will show
  that the coupled winds result in a less efficient feedback with
  respect to the original implementation.
\item {\bf ay37edw500}: Energy-driven winds of velocity $v_{\rm
  w}=500$ km s$^{-1}$ and Arimoto-Yoshii IMF.
\item {\bf sa37edw500}: Energy-driven winds of velocity $v_{\rm
  w}=500$ km s$^{-1}$ and Salpeter IMF (runs kr37edw500, ay37edw500
  and sa37edw500 are identical except for the IMF).
\item {\bf kr37p400edw500}: Energy-driven winds of velocity $v_{\rm
  w}=500$ km s$^{-1}$ and Kroupa IMF, with a number of particles (gas
  + dark matter) equal to $2\times400^{3}$. This simulation is
  identical to the kr37edw500 except for the larger number of
  particles and was run in order to check for resolution effects.
\end{itemize}
The name of each simulation reflects the choice of the IMF, the box
size \citep[even if the box size is equal to 37.5 $h^{-1}$ Mpc for all
  the simulations, we report it in order to be consistent with other
  works, e.g.][]{oppe06,oppe07} and the feedback prescription adopted
(only in the case of the resolution test run, kr37p400edw500, the
number of particles is specified in the name). In this paper, we label
as ``reference runs'' the first three of the list above: kr37edw500,
kr37mdw and kr37agn.

We use the {\small CLOUDY} code \citep{ferland} to compute
a-posteriori the CIV ionization fractions for each gas particle. As
for the work presented in \citet{tex09}, we choose the {\small HM05}
option in {\small CLOUDY}, which consists of a UVB made by QSOs and
galaxies with a 10\% photon escape fraction and which is in agreement
with other observational constraints (Bolton et al. 2005). In the left
panel of Figure \ref{fig:CIV_hm05new} we plot the CIV ionization
fraction, as a function of the gas overdensity $\delta_{\rm
  gas}+1=\rho_{\rm gas}/\bar{\rho}$, at $z=3$ for a gas temperature of
10$^{4.6}$ K (solid black line) and 10$^{5.0}$ K (dot-dashed green
line). In the right panel we plot the same ionization fraction, as a
function of the gas temperature $T_{\rm gas}$, at $z=3$ for an
overdensity of $\delta_{\rm gas}\sim 50$ (solid black line), this
value being representative of the outskirts of dark matter haloes, and
$\delta_{\rm gas}\sim 10$ (dot-dashed green line).

\begin{figure*}
\includegraphics[width=8.5cm]{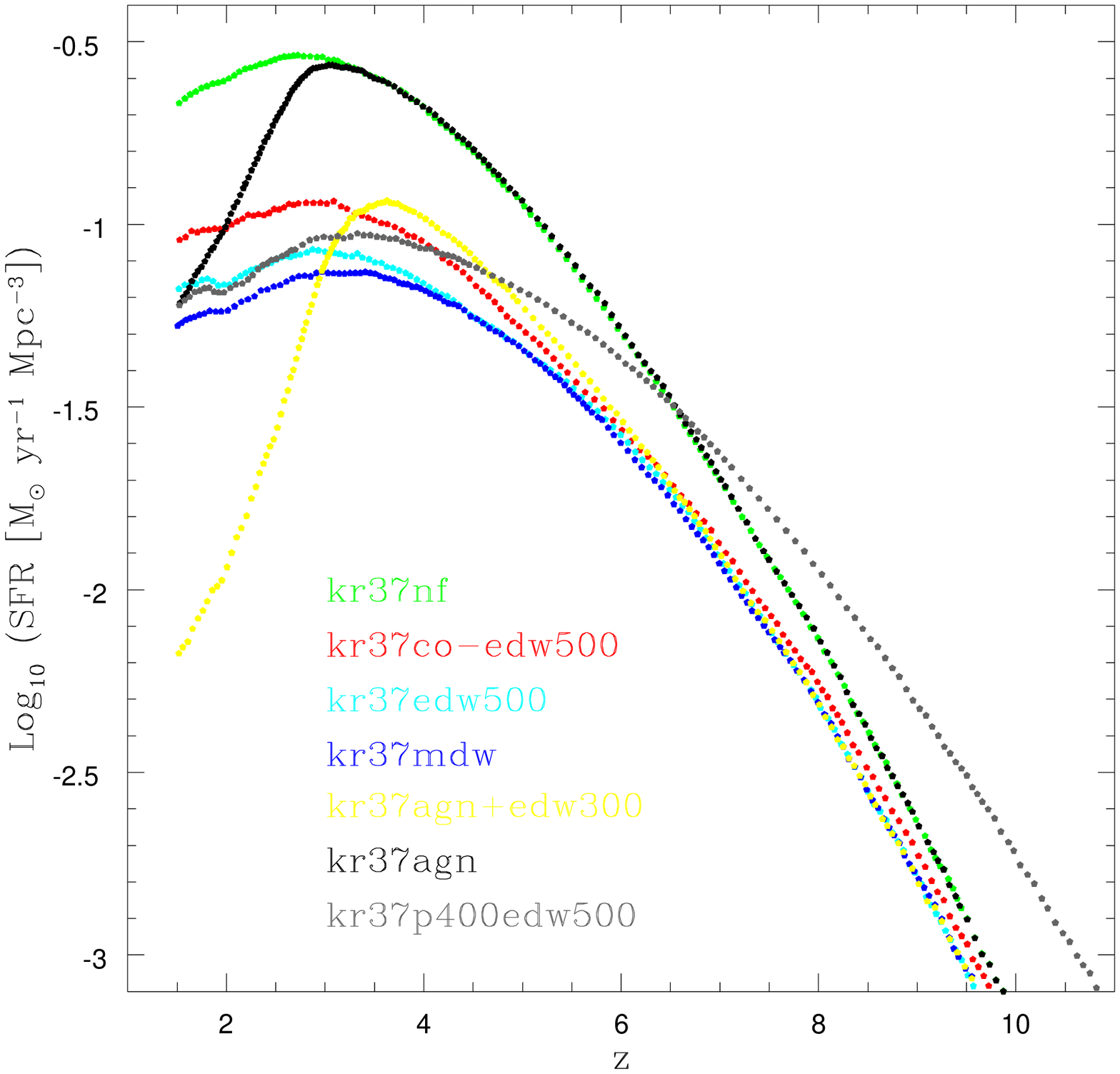}
\includegraphics[width=8.5cm]{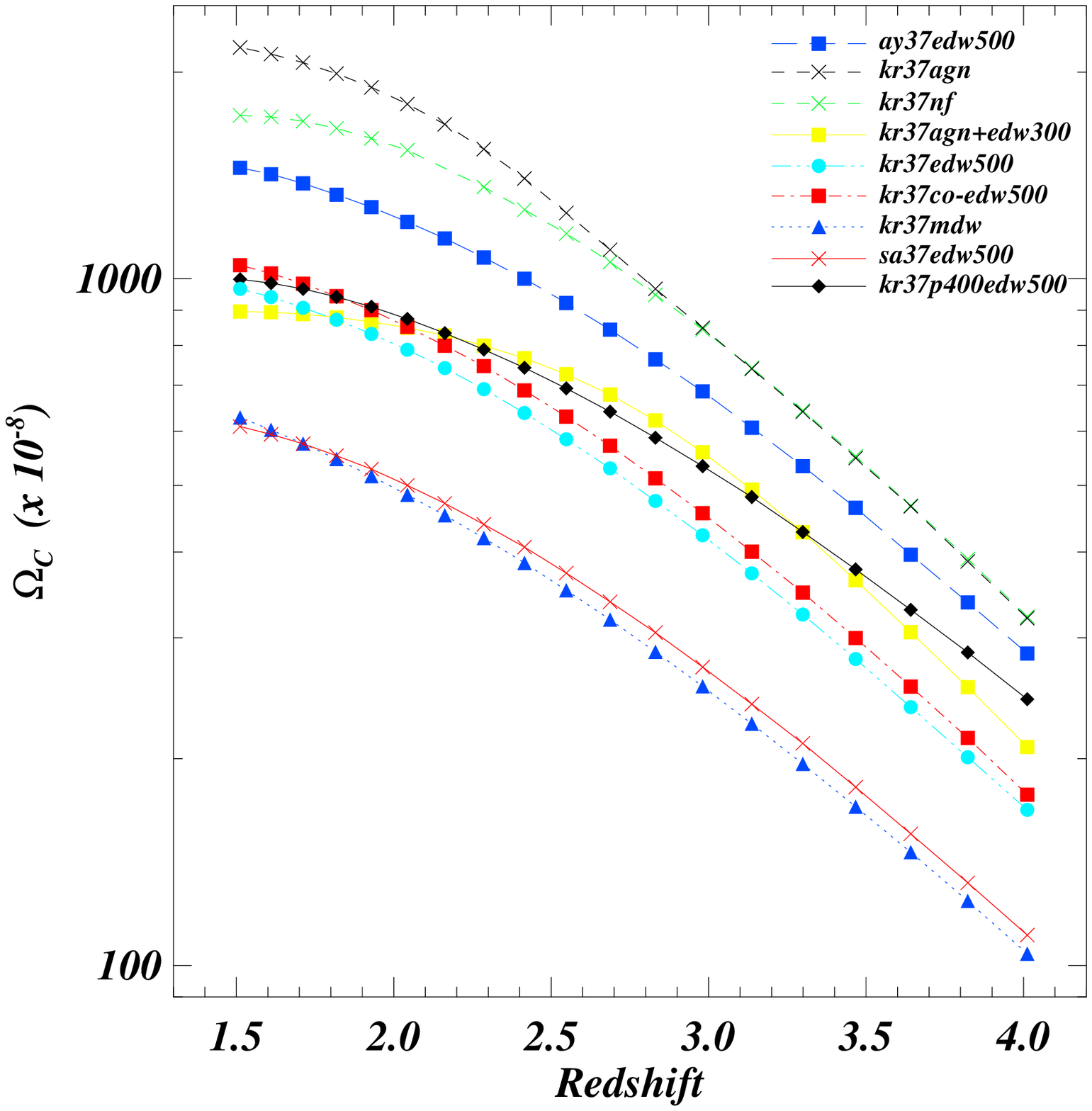}
\caption{{\it {Left Panel}}: cosmic star formation rate (SFR) for some
  of the hydrodynamic simulations of Table \ref{tab:sim_civ}. {\it
    Right Panel}: evolution of the total $\Omega_{\rm C}$ as a
  function of redshift for all the hydrodynamic simulations of Table
  \ref{tab:sim_civ}.}
\label{fig:om_civ}
\end{figure*}

\section{Star formation rates and evolution of the total CIV content}
\label{sec:sfrprop}

In this Section we analyse the cosmic star formation rates (SFRs) and
the evolution of the total Carbon content in the different
simulations.

In the left panel of Figure \ref{fig:om_civ} we report the total SFR
of the simulated volume as a function of redshift. All the
simulations, except the kr37p400edw500 run, have the same SFR down to
redshift $z\sim9.5$, when they start to differ due to their distinct
feedback recipes. Run kr37nf (no-feedback, green dots) has the highest
star formation rate compared to the other runs because there is no
effective mechanism able to quench the star formation. Run kr37agn
(AGN-feedback, black dots) follows the no-feedback run kr37nf down to
redshift $z=3$, but then the AGN feedback becomes effective and the
SFR suddenly decreases. Runs kr37co-edw500 (coupled energy-driven
winds, red dots), kr37edw500 (energy-driven winds, cyan dots) and
kr37mdw (momentum-driven winds, blue dots) show more or less the same
trend but for the kr37co-edw500 the SFR is slightly higher due to the
fact that the coupled winds are less efficient than the normal winds
in making the central region of the haloes devoid of gas and therefore in
suppressing the star formation. The kr37agn+edw300 simulation (AGN +
energy-driven winds, yellow dots) follows the other ``wind'' runs
(kr37co-edw500, kr37edw500 and kr37mdw) until redshift $z=3$ but then
the AGN feedback starts to become active and suppresses the star
formation very efficiently. Finally, at $z>8$, the kr37p400edw500 run
(resolution test, grey dots), shows a higher star formation with
respect to the other runs due to the improved resolution of this
simulation that can resolve higher densities at earlier times. Going
to lower redshift the SFR of the kr37p400edw500 agrees very well with
that of the kr37edw500 run, which has the same boxsize and parameters
but poorer resolution. This confirms the numerical convergence of our
simulations in the regime we are interested in. We stress that the aim
of this plot is to underline the effect of the different feedback
prescriptions on the SFR. For this reason observational data are not
plotted, because our simulations can be tuned in order to fit the
observed cosmic star formation rate at low redshift and, in this
sense, kr37edw500 is the reference simulation: all the other runs in
this plot differ only for the feedback prescription (all of them have
Kroupa IMF) and were not re-tuned. We also did not plot the two runs
with different IMFs ay37edw500 (Arimoto-Yoshii IMF) and sa37edw500
(Salpeter IMF), because we do not change the star formation efficiency
as it would be required in order to match the observables when the
number of massive stars per unit mass of formed stars changes: we are
more interested in the chemical and energetic effect of the IMF.

In the right panel of Figure \ref{fig:om_civ} we show the evolution
with redshift of the Carbon cosmological mass density $\Omega_{\rm
  C}(z)$, calculated considering the sum of the Carbon content
associated to each particle inside the cosmological box. Energy-driven
winds runs kr37edw500 (cyan circles$-$triple-dot-dashed line),
ay37edw500 (blue squares$-$dashed line) and sa37edw500 (red
crosses$-$solid line) are identical except for the IMF, therefore the
comparison of these three simulations is useful to understand how
Carbon production is affected by the choice of the IMF. The evolution
of $\Omega_{\rm C}$ is in fact the same, but with different
normalization: ay37edw500 has Arimoto-Yoshii IMF and produces more
Carbon than kr37edw500 run (Kroupa IMF) which itself produces more
Carbon than sa37edw500 (Salpeter IMF). With the Kroupa IMF, the
relative fraction of intermediate-mass stars, i.e. stars with masses
$2M_{\rm \odot}\leq m\leq 8M_{\rm \odot}$ , representing important
Carbon producers, is larger than with the Salpeter IMF
\citep{calu06,calu09}. For this reason, $\Omega_{\rm C}$ obtained with
the Kroupa IMF is higher than that obtained with the Salpeter IMF. On
the other hand, the assumption of a top-heavy IMF, such as the one of
\citet{ay87}, strongly enhances the number of massive stars but
suppresses the fraction of intermediate mass stars, with the overall
result of a higher Carbon density at any cosmic epoch. Interestingly,
run kr37mdw (momentum-driven winds, blue triangles$-$dotted line)
follows the sa37edw500 even if the latter produces much less CIV as we
will show in Section \ref{sec:omcivev} (Figure \ref{fig:omcivlines},
left panel). Run kr37co-edw500 (coupled energy-driven winds, red
squares$-$dot-dashed line) has the same trend as kr37edw500 but with a
slightly higher normalization. As we stated above, this is due to the
fact that the coupled winds are less efficient than the normal winds
and therefore this simulation has a higher star formation that results
in a higher Carbon abundance. This situation is even more extreme for
the kr37nf (no-feedback, green crosses$-$dashed line) and kr37agn
(AGN-feedback, black crosses$-$dashed line) runs, in fact down to
redshift $z\sim3$ both of these simulations do not suppress
effectively the star formation and for them $\Omega_{\rm C}$ is
higher. Going to lower redshift AGN feedback becomes active in
quenching the star formation and lowers the amount of gas used by the
stars and converts in other species (note the excess in $\Omega_{\rm
  C}$ with respect to the kr37nf run at $z<2.7$). The kr37agn+edw300
run (AGN + energy-driven winds, yellow squares$-$solid line) shows the
same $\Omega_{\rm C}$ trend of kr37edw500 even if with higher
normalization at high redshift due to the lower strength of the winds
(300 km s$^{-1}$ instead of 500 km s$^{-1}$). At low redshift AGN
feedback becomes active, but, differently from the kr37agn case: $i)$
winds have already made the haloes devoid of gas therefore reducing the
efficiency of the black holes accretion and thus the power of the AGN
feedback; $ii)$ winds suppress star formation at high redshift
lowering the Carbon content, so when the AGN starts to work the net
result is a further decrease of $\Omega_{\rm C}$. Finally, at high
redshift the kr37p400edw500 run (resolution test, black
diamonds$-$solid line), shows a higher $\Omega_{\rm C}$ with respect
to the kr37edw500 due to the improved resolution of this simulation
(as explained above), while at low redshift the Carbon content of the
kr37p400edw500 approaches that of the kr37edw500 run.

Comparing the $\Omega_{\rm C}$ evolution of our simulations with the
one showed in \citet[][Figure 12]{oppe06}, we found that in general
the normalization of our runs is higher, although a very precise
comparison is not possible given the different setup of the
simulations used, especially regarding the chemical evolution
model. However, for the kr37mdw run (and also for the sa37edw500 run)
the agreement is much better, the latter simulation producing
momentum-driven galactic winds and therefore being the yardstick for
the comparison with the \citet{oppe06} work.

\section{Statistics of HI}
\label{sec_histati}

Before focusing on the CIV evolution, we address some flux statistics
related to the properties and the evolution of neutral hydrogen in the
IGM. For all the neutral hydrogen statistics discussed in this paper,
we considered HI {\it lines} (while for the CIV we considered from
time to time lines or sytems of lines).

For each simulation performed we have extracted several physical
quantities interpolated along lines-of-sight (LOSs) through the
box. Given the positions, velocities, densities and temperatures of
all SPH particles at a given redshift, we compute spectra along a
given LOS through the box following the procedure of
\citet{theunsetal98}. We divide the sight line into $N=1024$ bins of
width $\Delta$ in distance $x$ along the sight line. For a bin $i$ at
position $x(i)$ we compute the density and the density weighted
temperature and velocity from:
\begin{eqnarray}
\rho_{\rm X}(j)   &=& a^3\,\sum_{\rm i} X(i) {\cal W}_{\rm ij},\\ 
\left(\rho T\right)_{\rm X}(j) &=& a^3\,\sum_{\rm i} X(i) T(i) {\cal W}_{\rm ij},\\
\left(\rho v\right)_{\rm X}(j) &=& a^3\,\sum_{\rm i} X(i) \{a\dot x(i)+\dot a [x(i)-x(j)]\} {\cal W}_{\rm ij},
\end{eqnarray}
where $a$ is the scale factor, $X(i)$ is the abundance of species $X$
of SPH particle $i$, assuming ionization equilibrium, and ${\cal
  W_{\rm ij}}=mW(q_{\rm ij})/h^3_{\rm ij}$ is the normalized SPH
kernel. Here $W$ is the SPH kernel, $m$ is the SPH particle mass and:
\begin{eqnarray}
q_{\rm ij} &=& \frac{a\left|x(i)-x(j)\right|}{h_{\rm ij}},\\
h_{\rm ij} &=& \frac{1}{2}\left[h(i)+h(j)\right],
\end{eqnarray}
with $h$ the physical softening scale. The optical depths of the
simulated QSO spectra are drawn in redshift space taking into account
the effect of the IGM peculiar velocities along the line-of-sight,
$v_{\rm pec,\parallel}$.

Taking the case of HI, even if the neutral hydrogen fraction $f_{\rm HI}$
is associated to each gas particle and is stored in each simulation
snapshot, we follow \cite{nagamine04}, as in \citet{tex09}, to assign
a-posteriori a new mass in neutral hydrogen to gas particles above the
density threshold for the onset of the star formation (set to
$\rho_{\rm th}=0.1$ cm$^{-3}$ in terms of the number density of
hydrogen atoms) which reads:
\begin{eqnarray}
m_{\rm HI}\,\,\,\,\, =&f_{\rm HI}\, m_{\rm H}\;\;\; & (\rho <
\rho_{\rm th})\\ m_{\rm HI}\,\,\,\,\, =&f_{\rm c}\, m_{\rm H}\;\;\;
&(\rho \geq \rho_{\rm th})\;,
\end{eqnarray}
with $f_{\rm HI}$ the neutral hydrogen fraction that depends on the
UVB used, $m_{\rm H}$ the hydrogen mass of the particle ($f_{\rm HI}$
and $m_{\rm H}$ are determined self-consistently inside the code),
$f_{\rm c}$ the fraction of mass in cold clouds and $\rho_{\rm th}$
the star formation threshold. Here $f_{\rm c}=\rho_{\rm c}/\rho$,
where $\rho_{\rm c}$ is the density of cold clouds and $\rho$ the
total (hot + cold) gas density. Individual molecular clouds cannot be
resolved at the resolution reachable in cosmological simulations, thus
$\rho_{\rm c}$ represents an average value computed over small regions
of the ISM \citep{springel2003}.

Our code follows self-consistently the evolution of Carbon (and of
other elements, see Section \ref{sec_sim}), then, using the {\small
  CLOUDY} code \citep{ferland}, we determine the ionization fractions
for the CIV ($\lambda\lambda$ 1548.204, 1550.781 \AA). The simulated
flux of a given ion transition at the redshift-space coordinate $u$
(in km s$^{-1}$) is $F(u)=\exp[-\tau(u)]$ with:
\begin{equation} 
\tau(u)={\sigma_{\rm 0,I} ~c\over H(z)} \int_{-\infty}^{\infty}n_{\rm I}(x) ~{\cal V}\left[u-x-v_{\rm pec,\parallel}^{\rm
    IGM}(x),\,b(x)\right]\ dx, 
 \label{eq1} 
\end{equation} 
where $n_{\rm I}$ is the ion number density, $\sigma_{\rm 0,I}$ is the
cross-section of the particular ion transition, $H(z)$ is the Hubble
constant at redshift $z$, $x$ is the real-space coordinate (in km
s$^{-1}$), $b=(2k_{\rm B}T/m_{\rm I}c^2)^{1/2}$ is the velocity
dispersion in units of $c$, ${\cal V}$ is the Voigt profile. These
spectra can be converted from velocity $v$ to observed wavelength
$\lambda$ using $\lambda=\lambda_{\rm 0}(1+z)(1+v/c)$.

\begin{figure*}
\centering
\includegraphics[width=6.5cm]{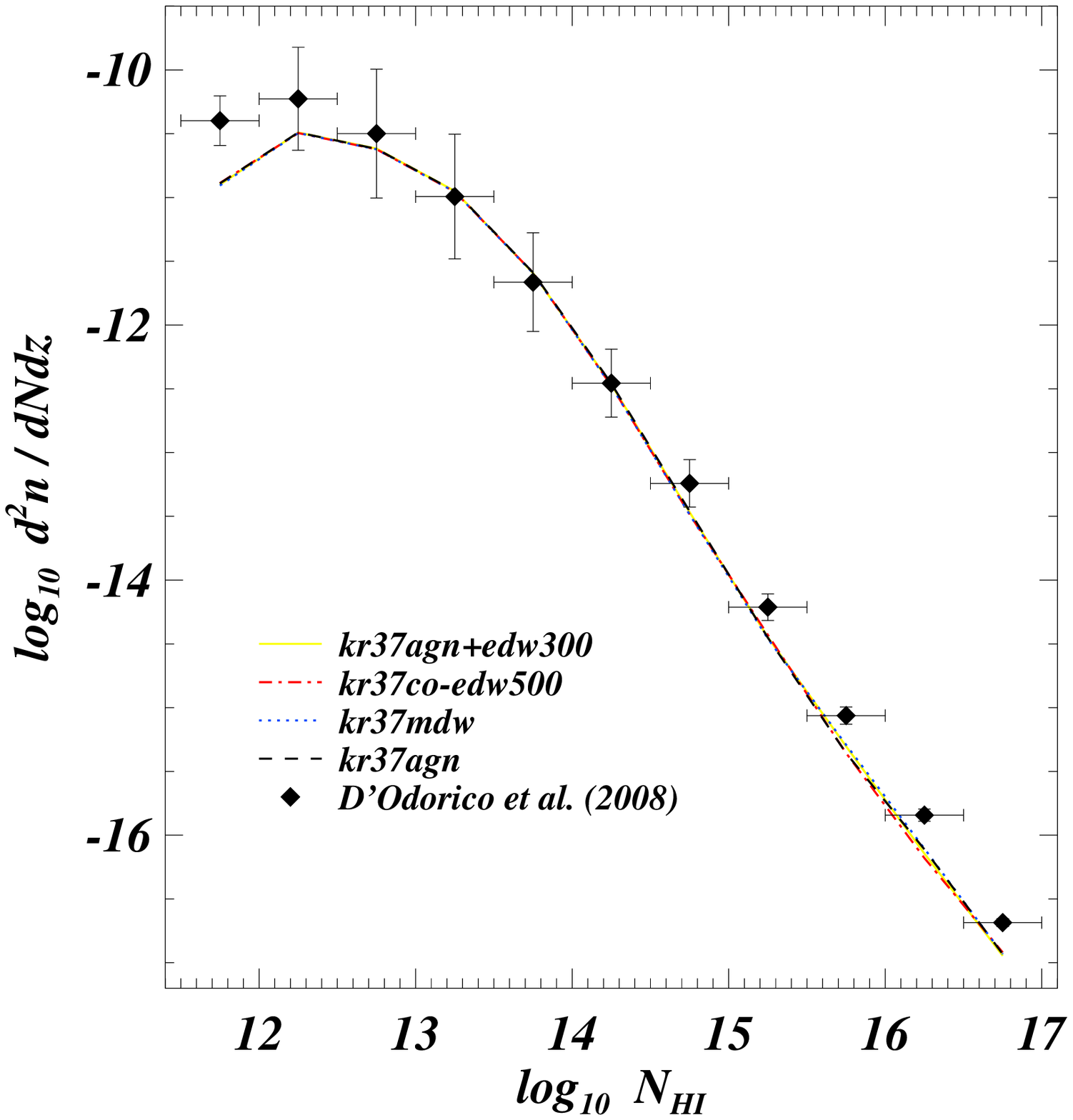}
\includegraphics[width=6.5cm]{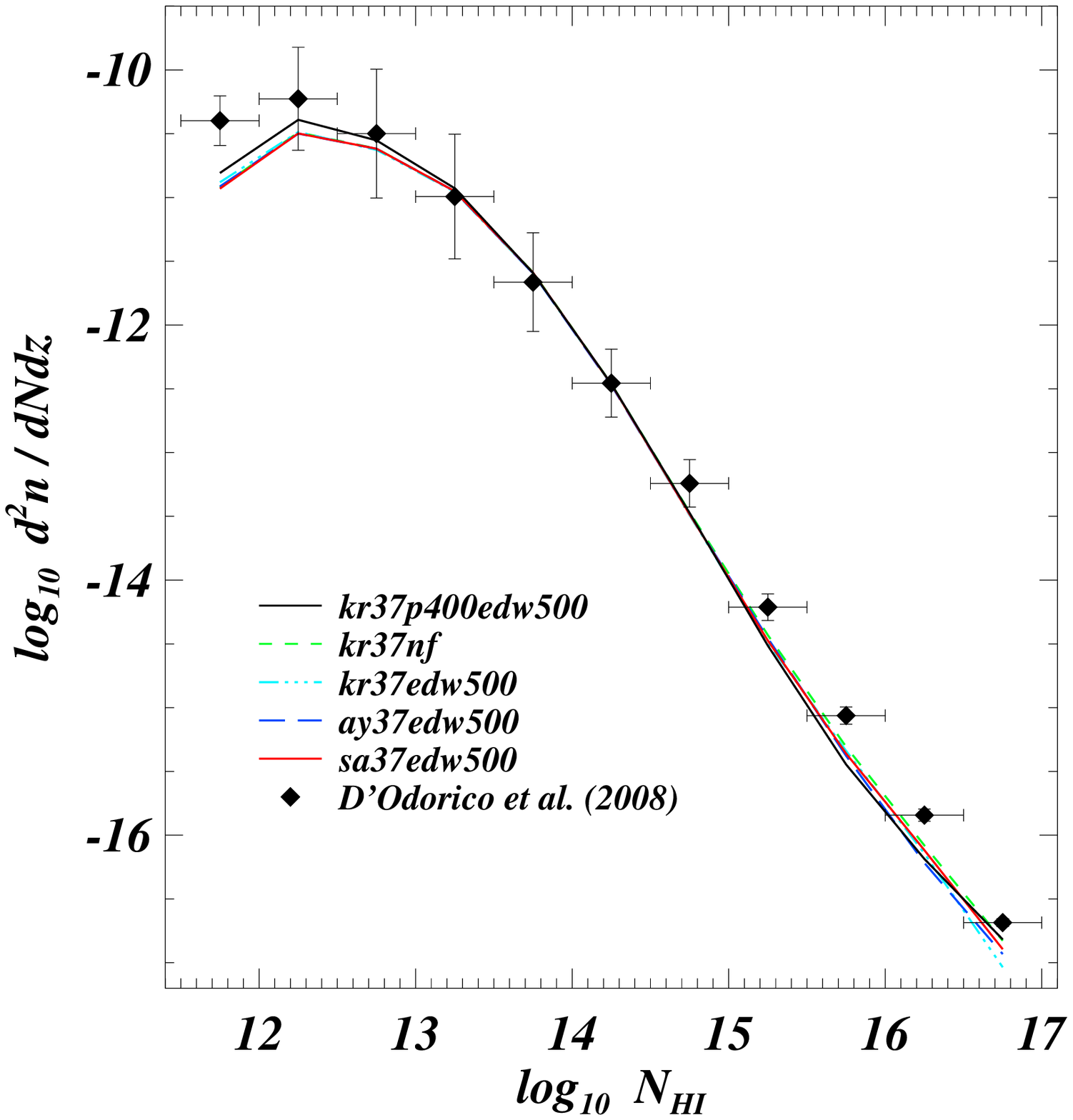}
\caption{Neutral Hydrogen (HI) column density distribution function at
  $z=3$. \textit{Left panel}: part I (first set of
  simulations). \textit{Right panel}: part II (second set of
  simulations). Data from \citet{vale08}.}
\label{fig:cddfh1z3}
\end{figure*}

\begin{figure*}
\centering
\includegraphics[width=6.5cm]{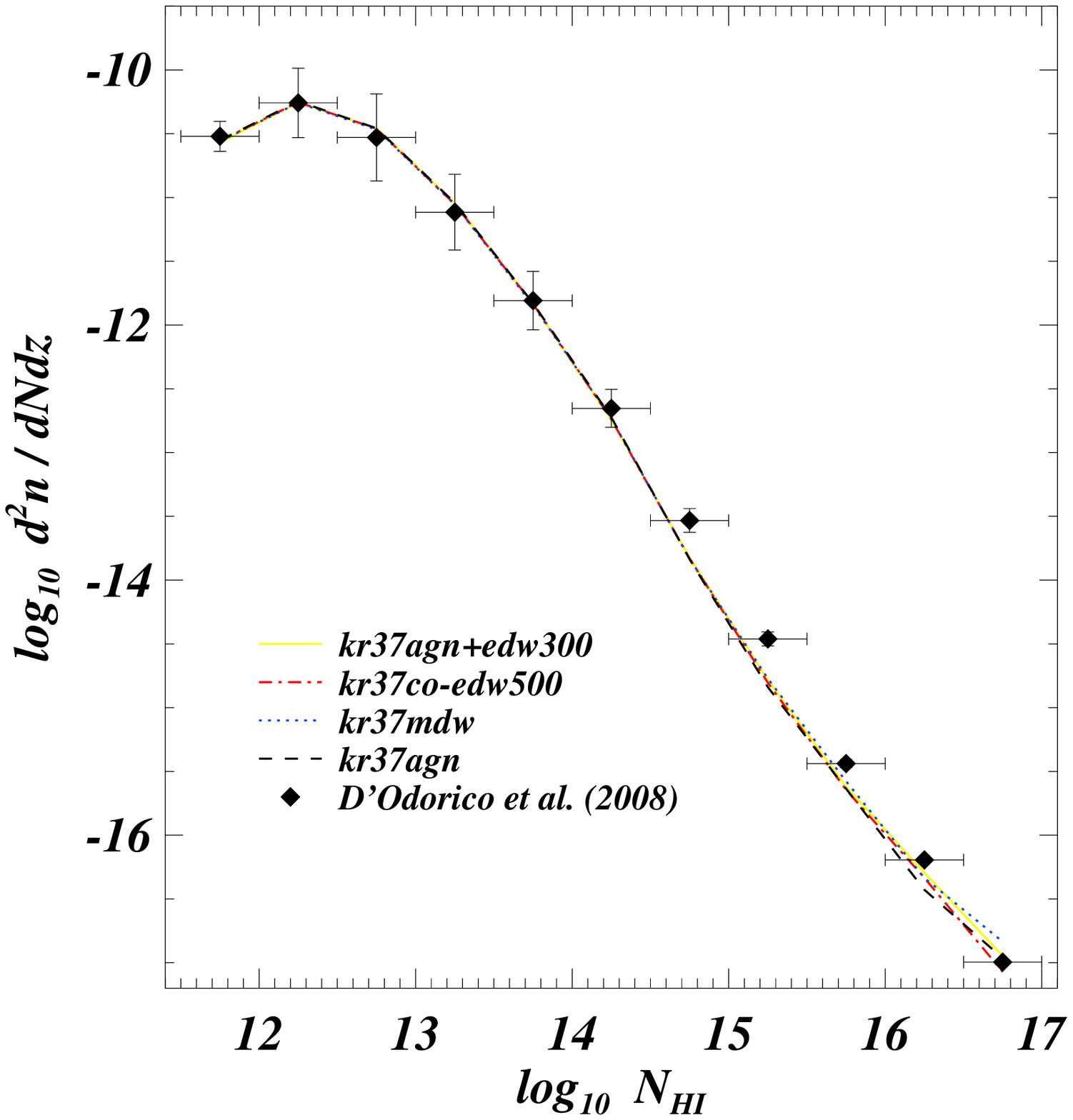}
\includegraphics[width=6.5cm]{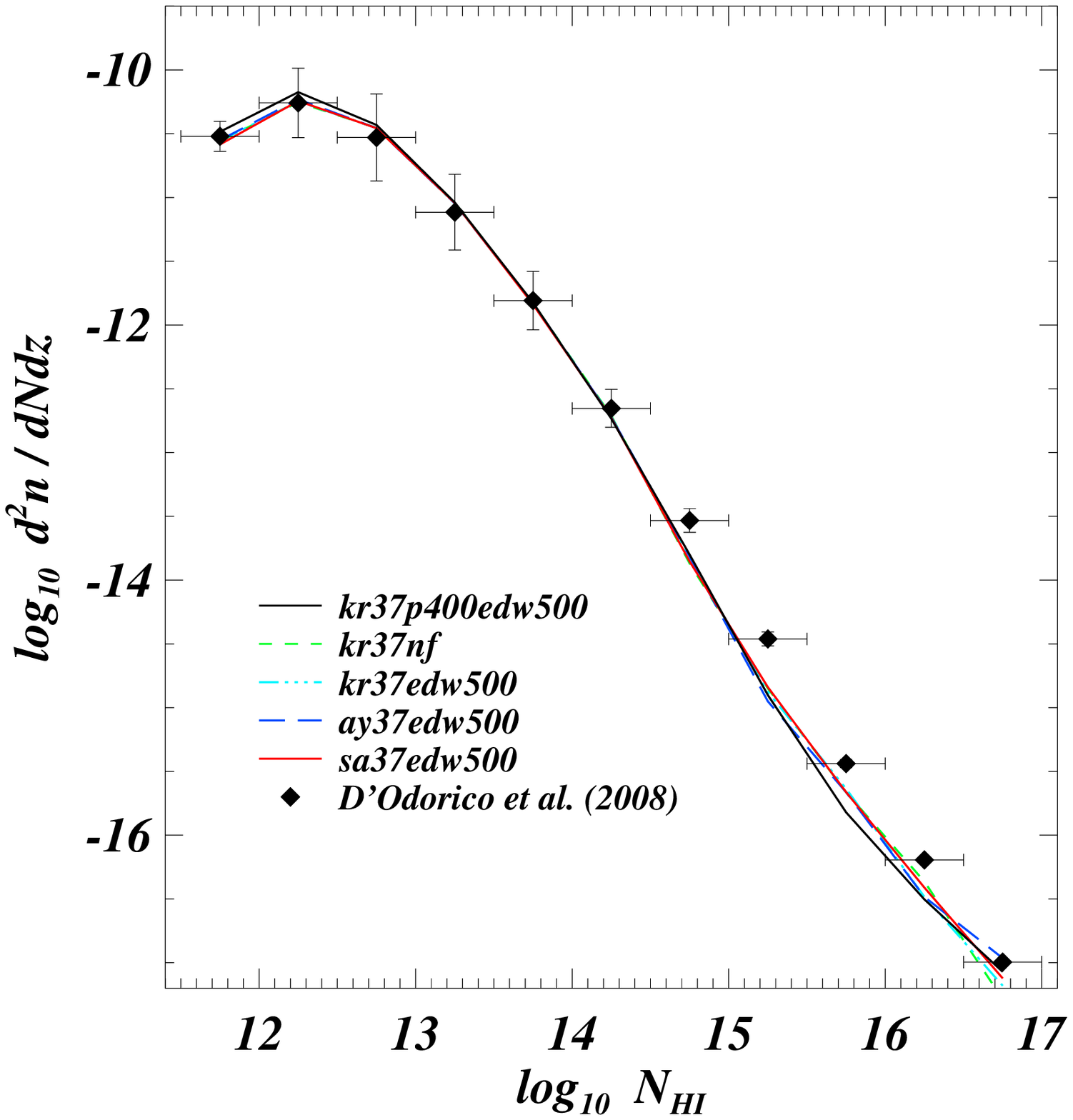}
\caption{As in Figure \ref{fig:cddfh1z3}, but at redshift $z=2.25$.}
\label{fig:cddfh1z2p25}
\end{figure*}

\begin{figure*}
\centering
\includegraphics[width=6.5cm]{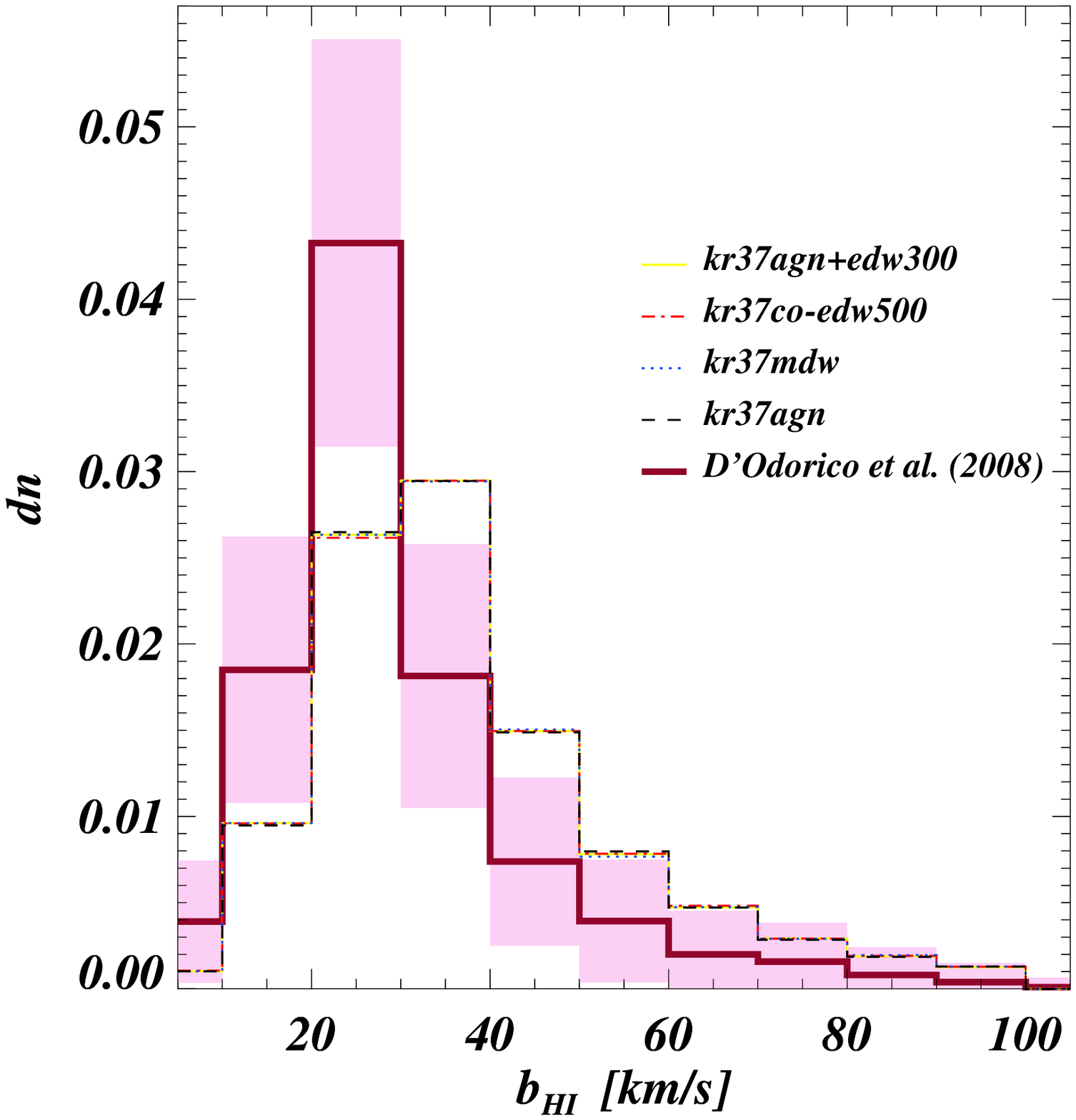}
\includegraphics[width=6.5cm]{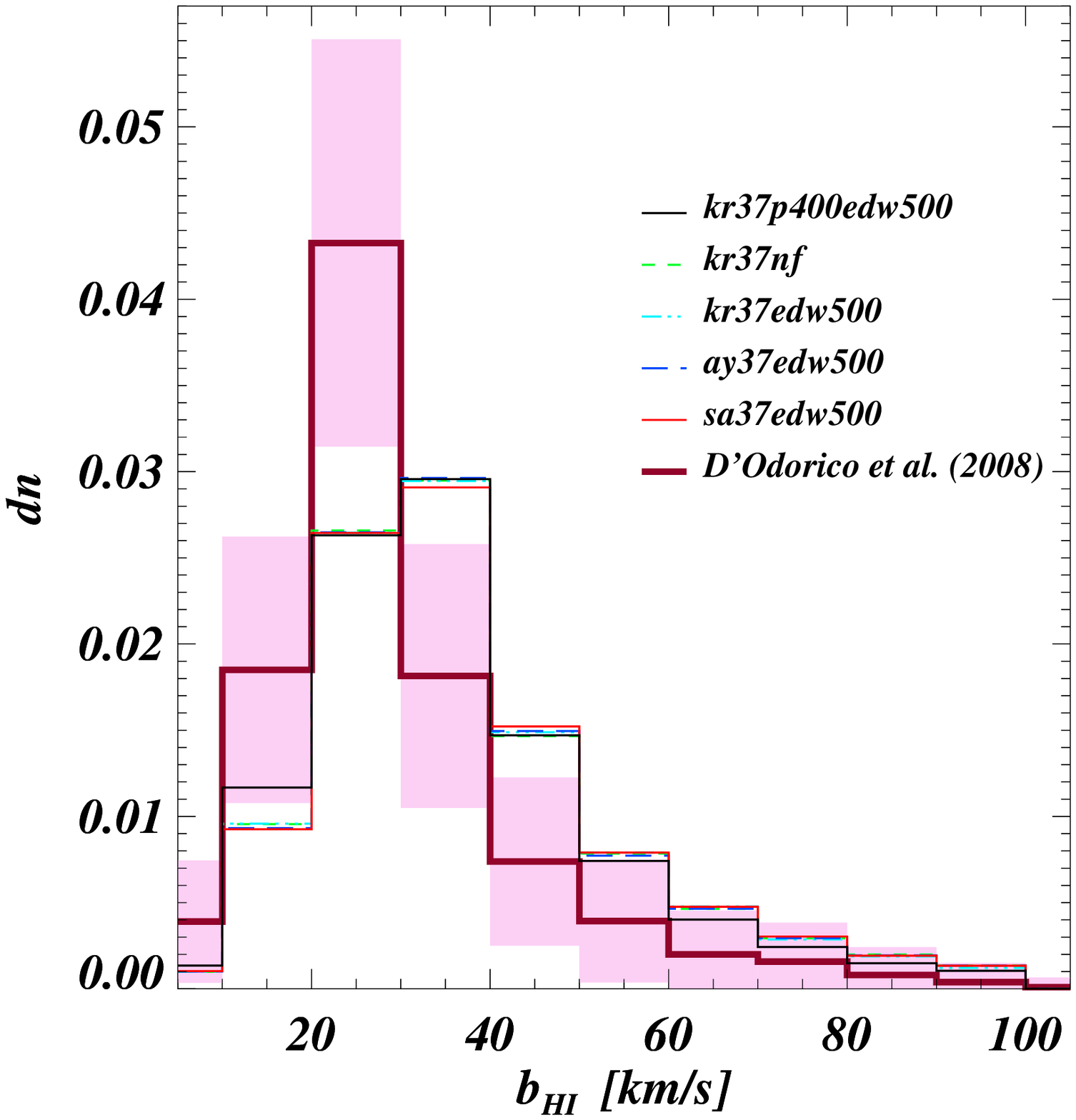}
\caption{Line-widths $b_{\rm HI}$ probability distribution function at
  $z=3$. \textit{Left panel}: part I. \textit{Right panel}: part
  II. In both panels, data from \citet{vale08} are showed by the
  purple solid line along with the associated Poissonian error (shaded
  region).}
\label{fig:bpdfh1z3}
\end{figure*}

\begin{figure*}
\centering
\includegraphics[width=6.5cm]{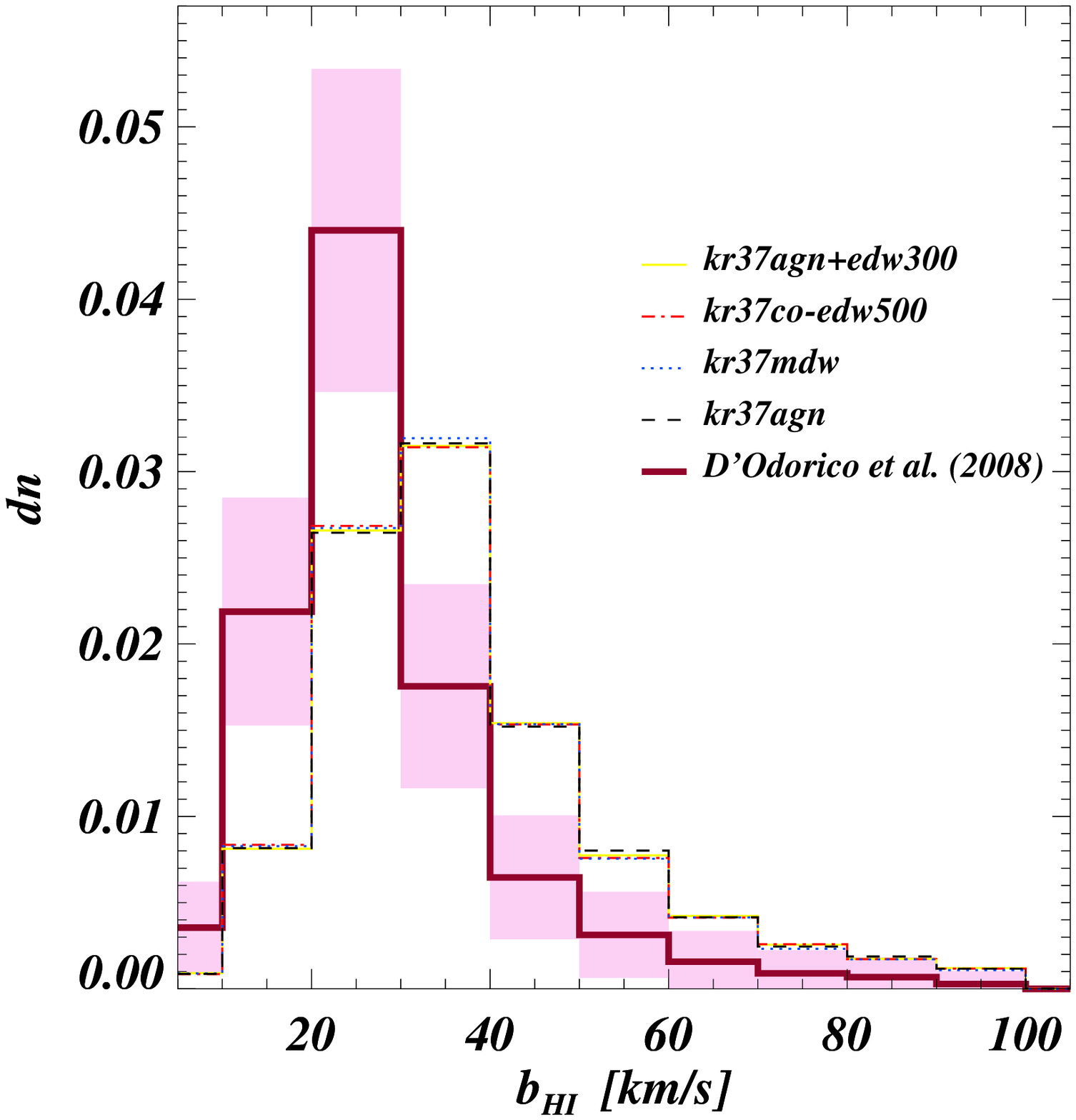}
\includegraphics[width=6.5cm]{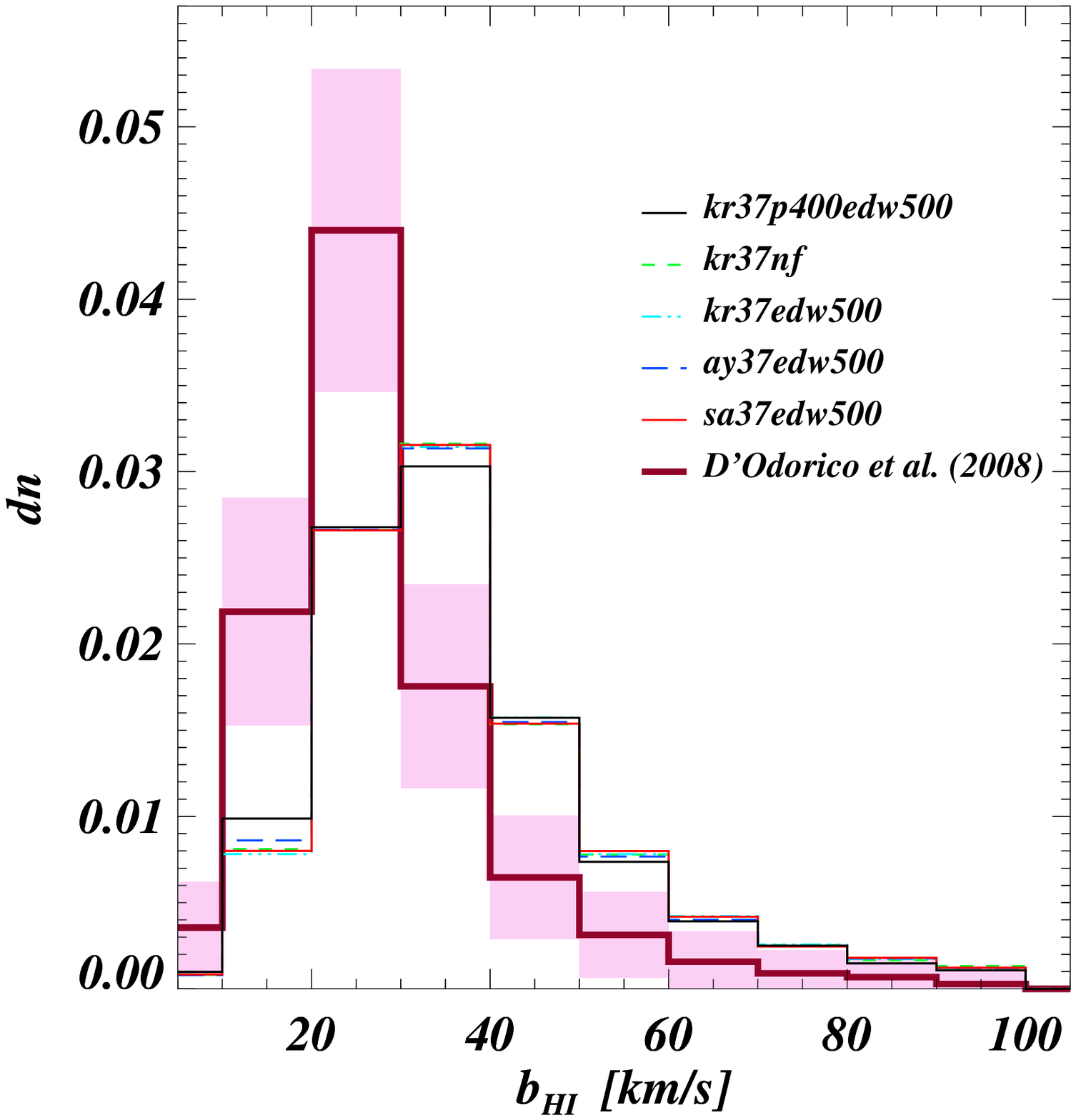}
\caption{As in Figure \ref{fig:bpdfh1z3}, but at redshift $z=2.25$.}
\label{fig:bpdfh1z2p25}
\end{figure*}

After having extracted the spectra along random LOS through the
cosmological box for a given simulation at a given redshift $z$, we
rescaled all the HI optical depths by a constant factor $A_{\rm HI}$
(ranging from $\sim0.43$ to $\sim0.73$, depending on redshift), in
such a way that their mean value was equal to the HI effective optical
depth, $\tau_{\rm HI}^{\rm eff}$, at that redshift, given by the
\citet{Kim2007} fitting formula:
\begin{equation}
\label{eq:eff_tau}
  \tau_{\rm HI}^{\rm eff}=(0.0023\pm 0.0007)(1+z)^{3.65\pm 0.21}.
\end{equation}
This rescaling ensures that our spectra match the observed mean
normalized flux of the Lyman-$\alpha$ forest at the appropriate
redshift: $\langle F\rangle_{\rm HI,obs}=\exp(-\tau_{\rm HI}^{\rm
  eff})$. We also rescaled all the CIV optical depths for the same
constant factor $A_{\rm HI}$. Then we convolved the spectra with a
Gaussian of 6.6 km s$^{-1}$ Full Width at Half-Maximum (FWHM) and we add noise in order to have
realistic spectra with signal-to-noise ratio $S/N=50$, to compare with
observations. Finally, we run the {\small VPFIT} code to fit each
spectra and recover HI and CIV lines' redshift, column density and
Doppler parameter. Among the lines fitted by {\small VPFIT} we have
considered only those with relative errors in the Doppler parameter
lower than 50\%.

\subsection{The HI column density distribution function}
\label{HICDDF}

In this Section we compute the Lyman-$\alpha$ forest column density
distribution function (HI CDDF) $d^2n/dNdz$, namely the number of
absorbers with HI column density in the range $[N,N+dN]$ and redshift
path in the interval $[z,z+dz]$. The redshift path at a given redshift
$z$ in simulated spectra is given by:
\begin{equation}
  dz=n_{\rm spec}(1+z)\frac{\Delta v}{c},
\end{equation}
where $c$ is the speed of light (in km s$^{-1}$) and $n_{\rm spec}$ is
the number of spectra taken at $z$ for a simulation with box size in
km s$^{-1}$ equal to $\Delta v$.

In Figure \ref{fig:cddfh1z3} we show the HI CDDF at redshift $z=3$ for
all the simulations of Table \ref{tab:sim_civ} splitted in two groups
for the sake of clarity. All the simulations fit very well the
observational data by \citet{vale08} down to column density $\log
N_{\rm HI} \,$(cm$^{-2}$)$\;=12.2$ and are all in agreement confirming
the fact that the HI column density distribution function is quite
insensitive to the different feedback prescriptions. Even though the
galaxies drive strong winds or AGN feedbacks, there is no discernible
effect on the Lyman-$\alpha$ forest. This is due to the fact that
winds or black hole ejecta expand preferentially into the lower
density regions and so keep the filaments that produce the hydrogen
lines intact \citep{theuns02}. The volume filling factor of the winds
is thus quite small and does not impact strongly on Lyman-$\alpha$
lines. In Figure \ref{fig:cddfh1z2p25} we show the HI CDDF at redshift
$z=2.25$: all the previous trends are confirmed and the agreement to
the data is very good also for $\log N_{\rm HI}\,$(cm$^{-2}$)$\;<12$.

\subsection{Probability distribution function of the HI Doppler parameter}
\label{sec:bHI_pdf}

In this Section we analyse the Lyman-$\alpha$ Doppler parameter,
$b_{\rm HI}$, probability distribution function. Throughout the rest
of the paper, both the HI and the CIV Doppler parameters are assumed
to be due only to the thermal width of the line and we neglected
possible turbulent contribution. Therefore, omitting the turbulent
contribution, the Doppler parameter basically measures the temperature
of the gas and is defined as:
\begin{equation}
  b_{\rm HI}=\sqrt{\frac{2k_{\rm B}T}{m_{\rm HI}}},
\end{equation}
where $m_{\rm HI}$ is the neutral hydrogen particle mass and $k_{\rm
  B}$ is the Boltzmann's constant.  In Figures \ref{fig:bpdfh1z3} and
\ref{fig:bpdfh1z2p25} we show the Lyman-$\alpha$ Doppler parameter
distribution function at redshifts $z=3$ and $z=2.25$, respectively,
for all the simulations of Table \ref{tab:sim_civ}, splitted in two
groups and compared with the data by \citet{vale08} (purple solid
line, while the shaded region represents the 1-$\sigma$ Poissonian
error). All runs are in good agreement with each other, while they
are shifted towards higher $b_{\rm HI}$ values with respect to the
observational data. At redshift $z=3$, the median value of the
observational Doppler parameters distribution is 25.9 km s$^{-1}$,
while all the simulations have median around $\sim34.5$ km
s$^{-1}$. At $z=2.25$, the data by \citet{vale08} have median value
equal to 24.8 km s$^{-1}$, while the simulations have median around
$\sim34.9$ km s$^{-1}$.

Our simulations produce HI gas that is too hot compared to the
observations. There can be many reasons for this, both physical and
numerical. Regarding the former, it is very important to stress the
effect of the UV background on the gas: if the UV background is too
strong the gas will be heated too much. As stated in Section
\ref{sec_sim}, we multiplied the \citet{HM96} heating rates by a
factor of 3.3, introduced to fit the \citet{schaye00b} constraints on the
temperature evolution of the intergalactic medium. It seems that this
correction does not work properly with new and higher quality data
sets.

However, minor part of the discrepancy is due to the lines fitting
procedure. When the absorption lines are weak and not well defined,
{\small VPFIT} can add relatively broad components in order to
minimize the $\chi^{2}$ statistics. This spurious components could
have large equivalent width and low optical depth, therefore they have
negligible impact on the column density distribution function because
of their relative small column density, but they can bias the Doppler
parameter probability distribution function. We will show later in
Section \ref{sec:bcivNciv} that this numerical effect is more evident
for the CIV statistics and how it is important in explaining some
features of the $b_{\rm CIV}-N_{\rm CIV}$ relation.

\begin{figure*}
\centering
\includegraphics[scale=0.53]{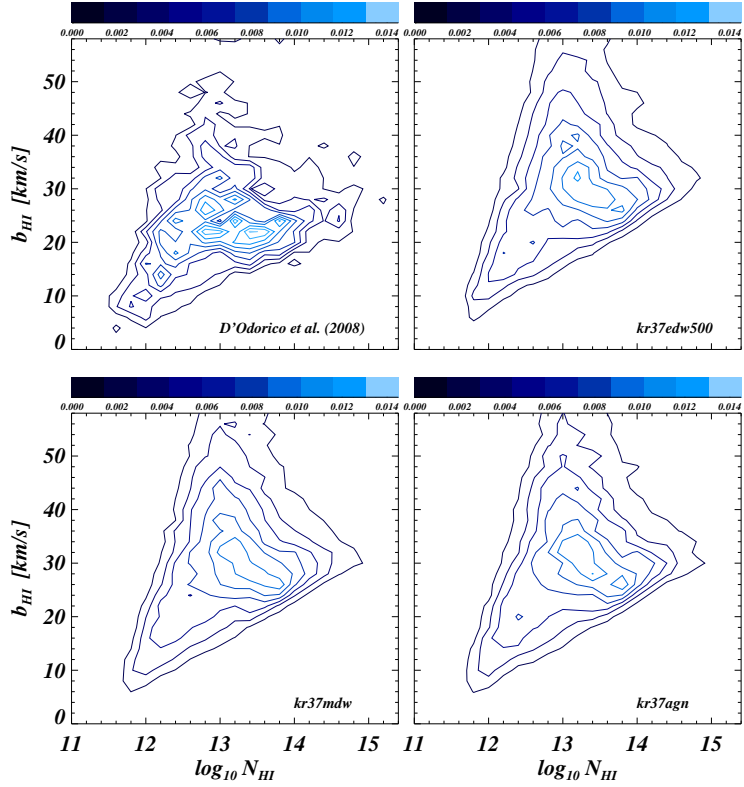}
\caption{$b_{\rm HI}-N_{\rm HI}$ relation at $z=3$. The horizontal
  bars are colour coded according to the fraction of points that fall
  in each bin with coordinate ($b_{\rm HI}$,$N_{\rm HI}$). Upper left
  panel: observational data from \citet{vale08}.}
\label{fig:bNh1_z3}
\end{figure*}

\begin{figure*}
\centering
\includegraphics[scale=0.53]{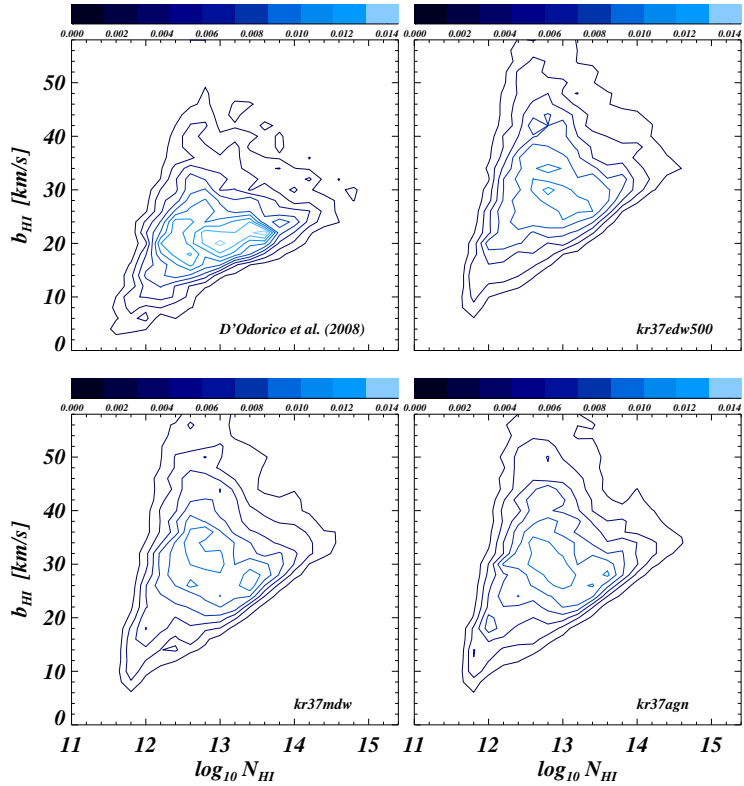}
\caption{As in Figure \ref{fig:bNh1_z3}, but at redshift $z=2.25$.}
\label{fig:bNh1_z2p25}
\end{figure*}

\subsection{Column density-Doppler parameter relation}
\label{sec_bHINHI}

In Figures \ref{fig:bNh1_z3} and \ref{fig:bNh1_z2p25} we plot the
$b_{\rm HI}-N_{\rm HI}$ relation, at redshift $z=3$ and $z=2.25$
respectively, for the reference simulations kr37edw500, kr37mdw and
kr37agn. The contour plots are colour coded according to the fraction
of points that fall in each bin with coordinate ($b_{\rm HI}$,$N_{\rm
  HI}$). In the upper left panels the observational data of
\citet{vale08} are reported. Both at redshifts $z=3$ and $z=2.25$ all
runs are in good agreement with each other and reproduce well the
observations, even if the simulations present Doppler parameters
slightly higher than the observational data (as we mentioned in the
previous Section). The mean slope of the data envelope in the column
density range $11.5<\log N_{\rm HI}\,$(cm$^{-2}$)$\;<15$ measures the
slope of the $T-\rho$ relation for the gas \citep{schaye2001}, and all
the runs show roughly the same right slope of the data at the two
redshifts.\\

To sum up, the fact that for our simulations all the different HI
statistics are in general agreement with the observational data,
confirms that we are catching the physics of the gas traced by the
neutral hydrogen. The discrepancies are due, mainly, to the choice of
the UV background \citep[assumed to be produced by quasars and
  galaxies as given by][with the heating rates multiplied by a factor of
  $3.3$, see Sections \ref{sec_sim} and \ref{sec:bHI_pdf}]{HM96} and,
in minor part, to a numerical effect introduced by {\small VPFIT} that
affects our analysis of the absorption lines in quasar spectra. We
will discuss in more detail the impact of this effect in the next
sessions regarding the CIV absorption lines statistics.

The resolution-test simulation kr37p400edw500 follows the other runs
in all the previous statistics, again confirming the numerical
convergence of our simulations for the HI evolution.

\section{The CIV column density distribution function}
\label{sec_civcddf}

\begin{figure*}
\centering
\includegraphics[width=6.5cm]{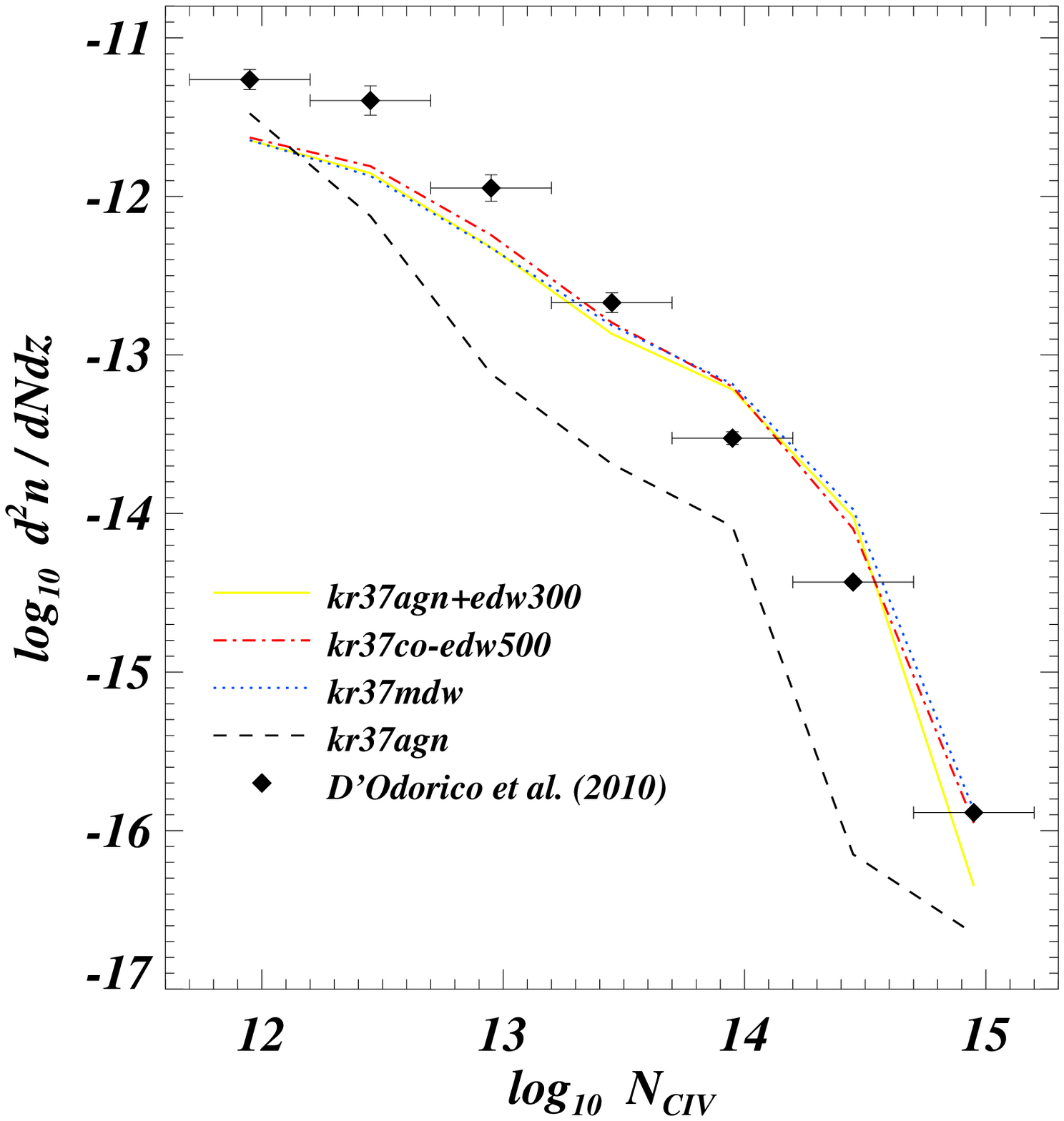}
\includegraphics[width=6.5cm]{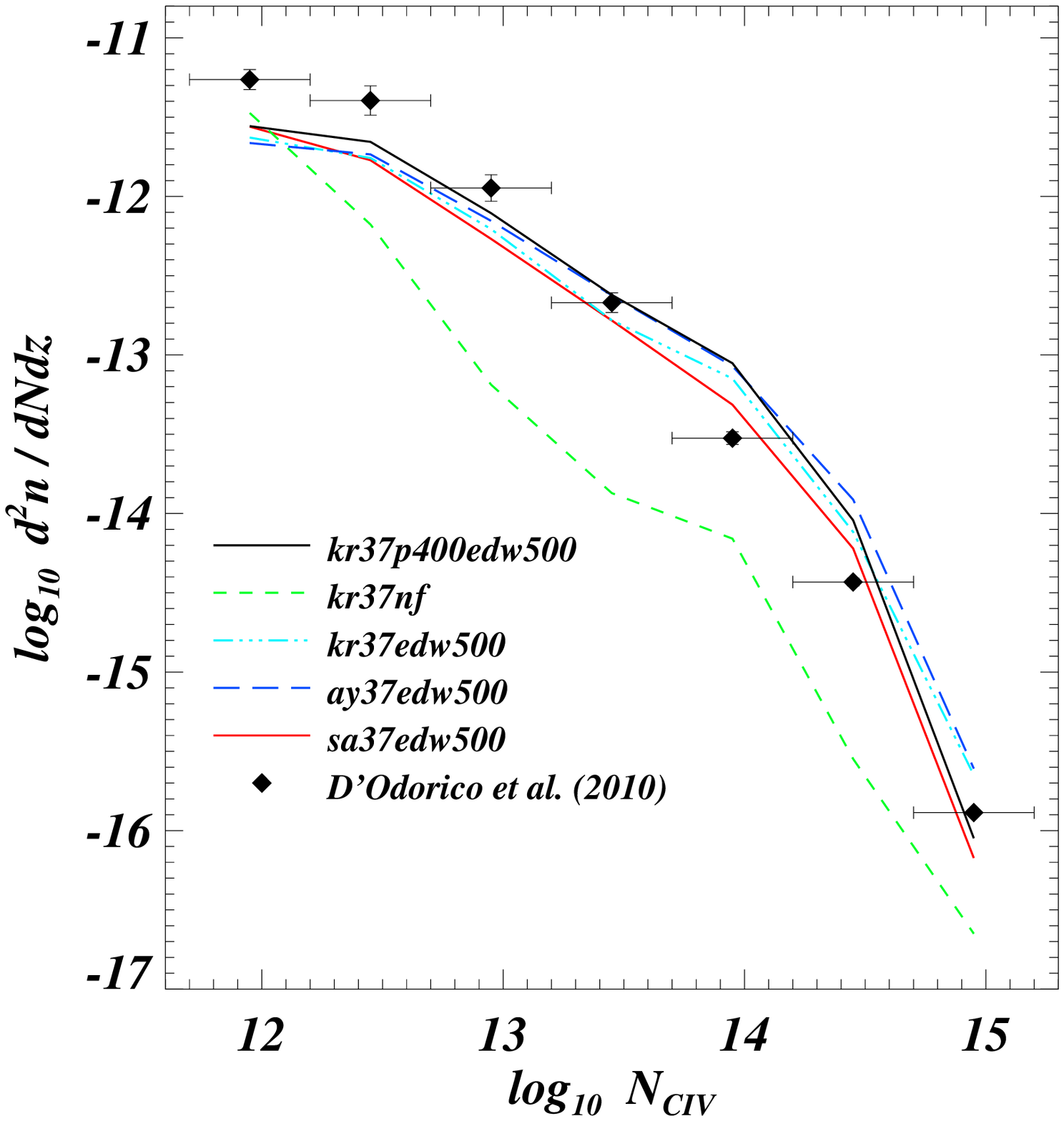}
\caption{CIV column density distribution function at
  $z=3$. \textit{Left panel}: part I. \textit{Right panel}: part
  II. Data from \citet{vale10}.}
\label{fig:cddfz3}
\end{figure*}

\begin{figure*}
\centering
\includegraphics[width=6.5cm]{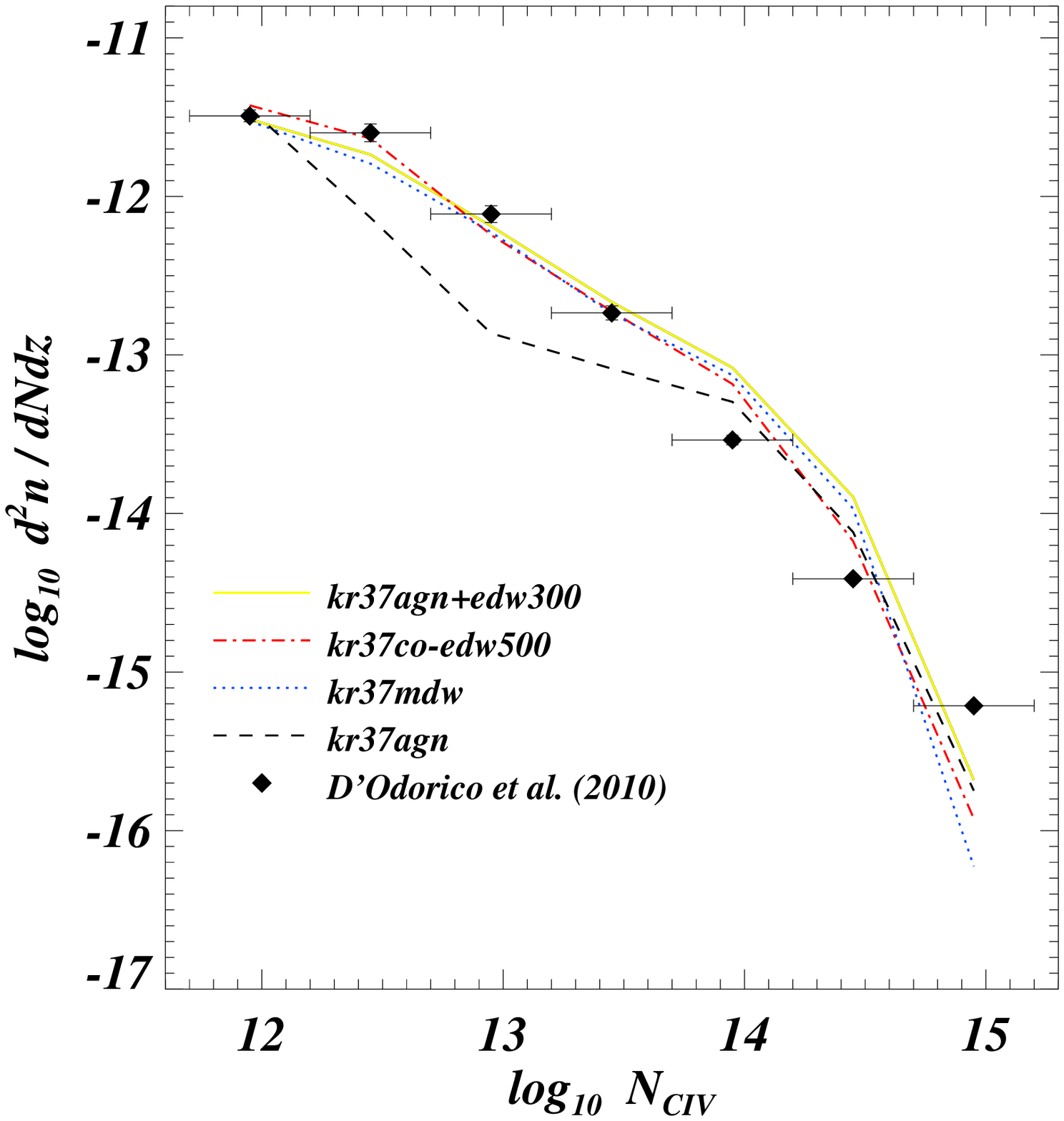}
\includegraphics[width=6.5cm]{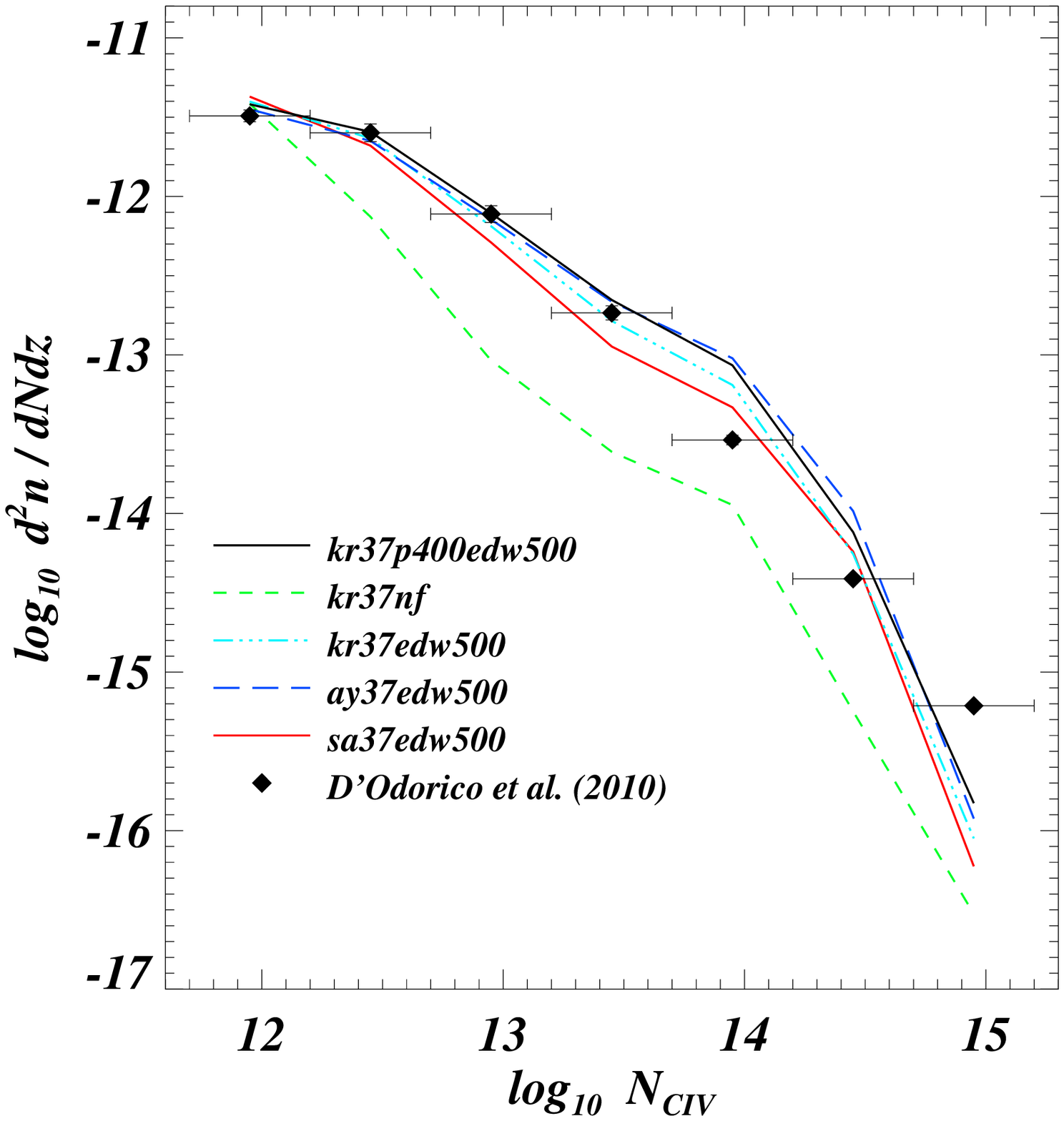}
\caption{As in Figure \ref{fig:cddfz3}, but at redshift $z=2.25$.}
\label{fig:cddfz2p25}
\end{figure*}

\begin{figure*}
\centering
\includegraphics[width=6.5cm]{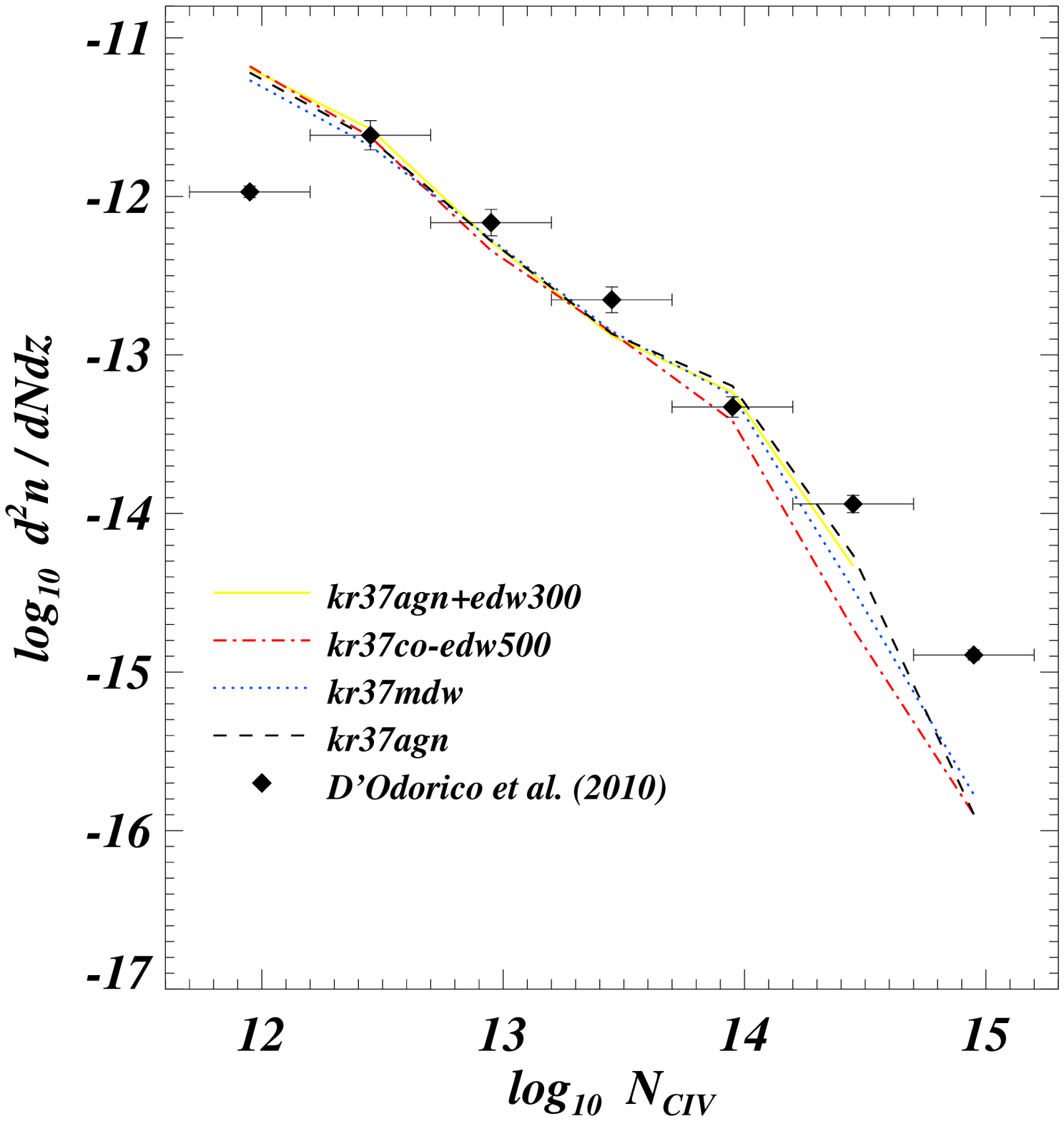}
\includegraphics[width=6.5cm]{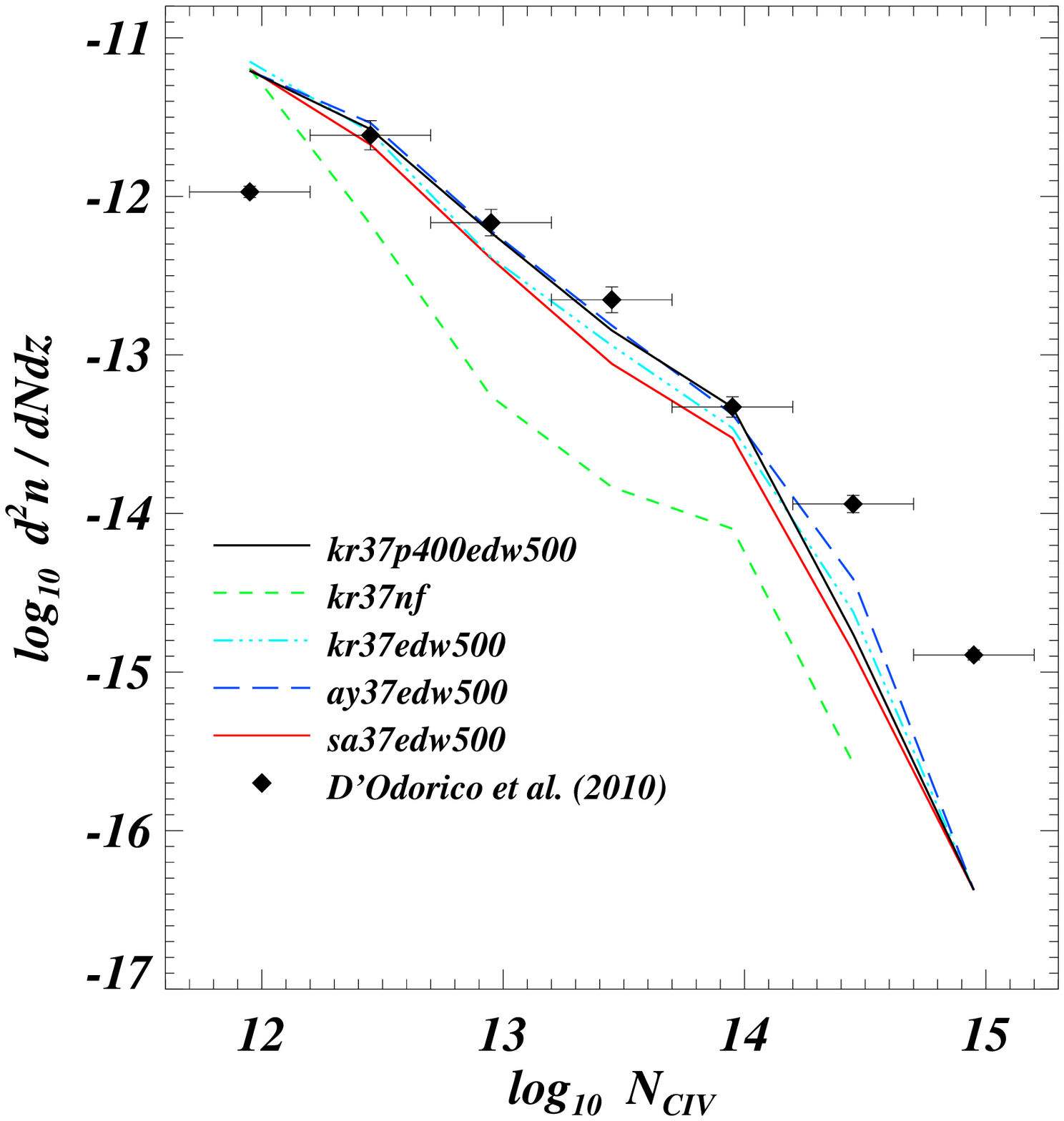}
\caption{As in Figure \ref{fig:cddfz3}, but at redshift $z=1.5$}
\label{fig:cddfz1p5}
\end{figure*}

In this Section we investigate the CIV column density distribution
function (CIV CDDF, already defined in Section \ref{HICDDF}), plotted
at redshift $z=3$, 2.25 and 1.5, respectively and along with the
\citet{vale10} data, in Figures \ref{fig:cddfz3}, \ref{fig:cddfz2p25}
and \ref{fig:cddfz1p5}. We considered for this analysis {\it systems}
of lines as defined in Section \ref{obdatasamp}.

At $z=3$ (Figure \ref{fig:cddfz3}), runs (left panel) kr37agn+edw300
(AGN + energy-driven winds), kr37co-edw500 (coupled energy-driven
winds), kr37mdw (momentum-driven winds) and (right panel) kr37edw500
(energy-driven winds, Kroupa IMF), ay37edw500 (energy-driven winds,
Arimoto-Yoshii IMF), sa37edw500 (energy-driven winds, Salpeter IMF),
are in agreement with each other and reproduce well the observed
distribution down to $\log N_{\rm CIV}\,$(cm$^{-2}$)$\;\sim12.7$. In
contrast, for the simulations kr37nf and kr37agn the agreement is not
good. Simulation kr37nf (Figure \ref{fig:cddfz3} right panel, green
dashed line) was run, as a test, without any winds or AGN feedback: so
there is no effective mechanism to spread the enriched gas outside the
haloes. The result is that all the gas is trapped inside high density
haloes and most of the Carbon is locked back into stars. In the run
kr37agn (Figure \ref{fig:cddfz3} left panel, black dashed line) the
feedback mechanism is the AGN feedback, associated to the energy
released by gas accretion onto super-massive black holes. Comparing
the resulting distribution with that of kr37nf, we note that they are very
similar. The reason is that at redshift $z=3$ the AGN feedback is not
active yet, but it is just starting to work, so at high redshift this
simulation virtually behaves like the no-feedback one. Interestingly,
run kr37agn+edw300 (AGN + energy-driven winds) follows all the other
``wind'' runs even if in this case we have the combined effect of
winds and AGN feedbacks: this is due to the fact that winds start to
be effective at higher redshift than the AGN feedback, making the haloes devoid of gas and in this way reducing the efficiency of the black
holes accretion and the power of the AGN feedback.

\begin{figure*}
\centering
\includegraphics[width=8.5cm]{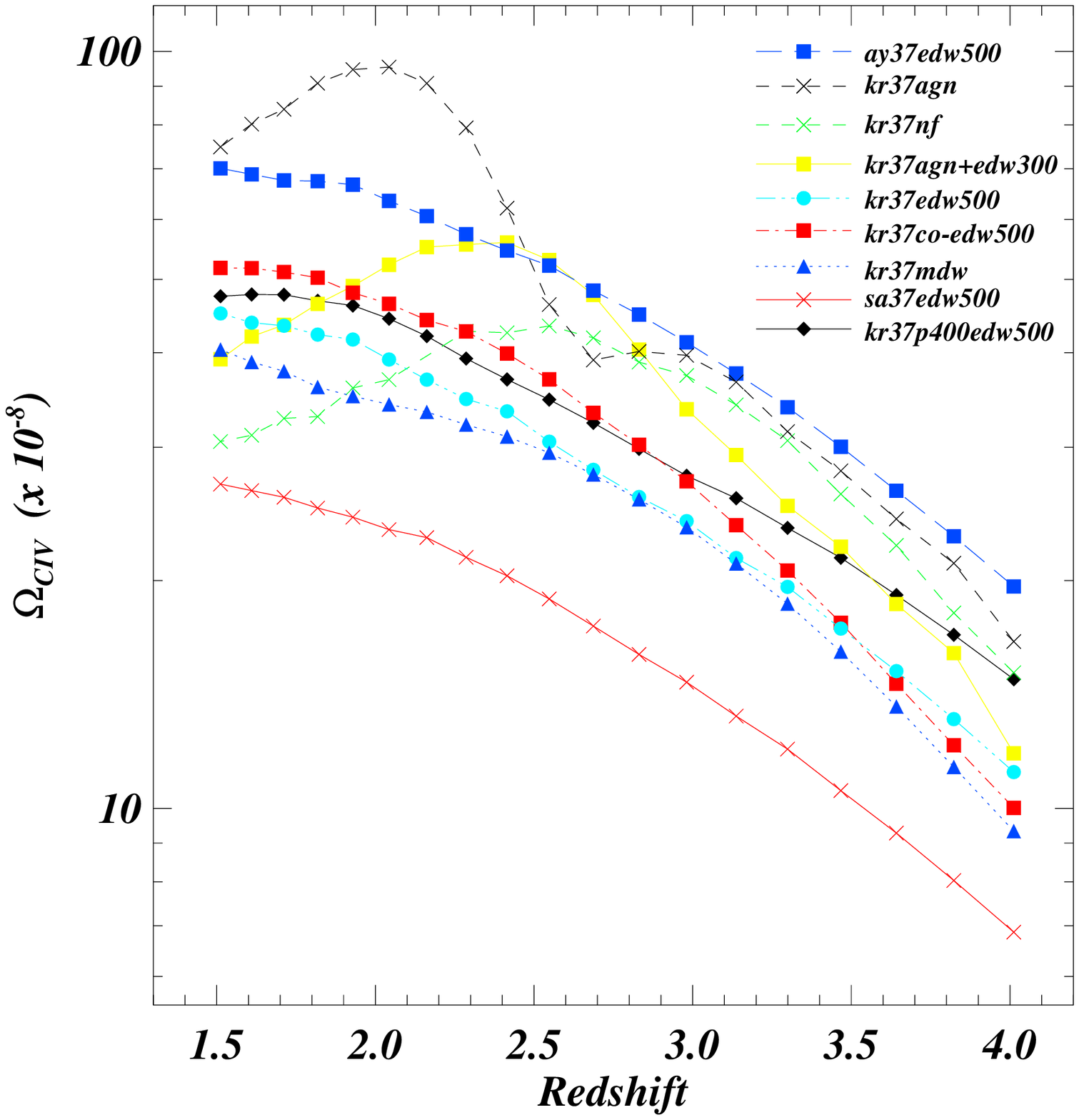}
\includegraphics[width=8.5cm]{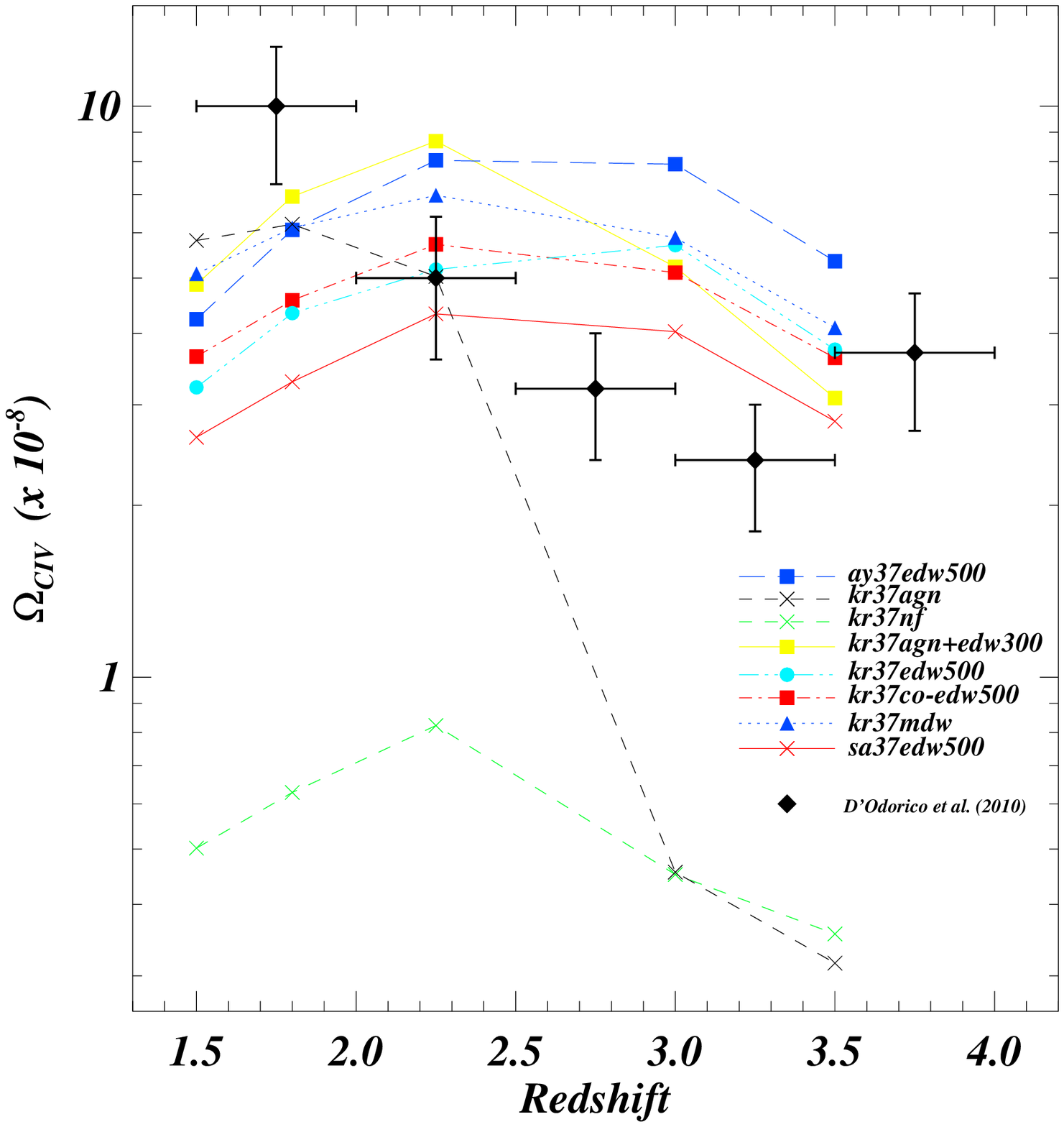}
\caption{{\it {Left Panel}}: evolution of the total $\Omega_{\rm CIV}$
  as a function of redshift for all the hydrodynamic simulations of
  Table \ref{tab:sim_civ}. {\it Right Panel}: $\Omega_{\rm CIV}$
  evolution with redshift, calculated considering the absorption lines
  in 1000 lines-of-sight through each box, for the different
  simulation boxes. The overplotted black diamonds show the
  observational data of \citet{vale10}.}
\label{fig:omcivlines}
\end{figure*}

At redshift $z=2.25$ (Figure \ref{fig:cddfz2p25}), the agreement with
observational data is very good down to $\log N_{\rm
  CIV}\,$(cm$^{-2}$)$\;=12$. The kr37nf run (no-feedback, right panel
dashed green line) suffers of the same problems described before,
while for the kr37agn run (AGN feedback, left panel dashed black line)
the discrepancies with the observational data decreased: a first hint
that going to lower redshift AGN feedback becomes more and more
efficient and the behaviour of this simulation starts to resemble that
of the ``wind'' runs.

At redshift $z=1.5$ (Figure \ref{fig:cddfz1p5}) all the simulations
produce too many systems with low column density and systematically fall slightly short of reproducing the distribution at $\log N_{\rm
  CIV}\,$(cm$^{-2}$)$\;>14.5$. Now the kr37agn run (AGN feedback, left
panel dashed black line) closely follows all the other runs,
confirming that at this low redshift AGN feedback is active and
powerful, while kr37nf run (no feedback, right panel dashed green
line) still fails in reproducing the correct trend. For this
statistics, at the three redshift considered, the resolution-test
simulation kr37p400edw500 follows the reference runs.

Note that at redshift $z=2.25$ we fit very well the observational data
in the low column density tail of the distribution, $\log N_{\rm
  CIV}\,$(cm$^{-2}$)$\;<12.2$, while at $z=3$ and $z=1.5$ the
agreement in this column density range gets worse.

\section{The evolution of the CIV cosmological mass density}
\label{sec:omcivev}

In the left panel of Figure \ref{fig:omcivlines} we show $\Omega_{\rm
  CIV}(z)$, calculated considering the sum of the CIV content
associated to each particle inside the cosmological box. Energy-driven
winds runs kr37edw500 (cyan circles$-$triple dot-dashed line),
ay37edw500 (blue squares$-$dashed line) and sa37edw500 (red
crosses$-$solid line) behave exactly as we described in Section
\ref{sec:sfrprop}: they differ only for the IMF and the ay37edw500
produces much more CIV than the other two, having Arimoto-Yoshii
IMF. Runs kr37co-edw500 (coupled energy-driven winds, red
squares$-$dot-dashed line) and kr37mdw (momentum-driven winds, blue
triangles$-$dotted line) follow almost the same trend as kr37edw500,
which has the same IMF of the two runs mentioned above. For the kr37nf simulation
(no-feedback, green crosses$-$dashed line), $\Omega_{\rm CIV}$ starts
to decrease at redshift $z\sim 2.7$. It is interesting to note that
for the simulation kr37agn (AGN-feedback, black crosses$-$dashed line)
$\Omega_{\rm CIV}$ suddenly increases after a small decrement at
redshift $z\sim 2.8$ (down to this redshift the kr37agn run follows
the no-feedback run). This is due to the suppression of star formation
produced by the powerful AGN feedback. Then, moving to lower redshift
the CIV cosmological density starts to approach all the other
runs. The same effect is visible in the kr37agn+edw300 run (AGN +
energy-driven winds, yellow squares$-$solid line). In this case the
effect is weaker because, as we explained in Section
\ref{sec_civcddf}, winds make the haloes devoid of gas therefore reducing the
efficiency of the black holes accretion and thus the power of the AGN
feedback. As for the total $\Omega_{\rm C}$ plot in the right panel of
Figure \ref{fig:om_civ}, at high redshift, the resolution-test
simulation kr37p400edw500, shows an higher $\Omega_{\rm CIV}$ with
respect to the kr37edw500, while the two runs are in agreement at low
redshift.

In the right panel of Figure \ref{fig:omcivlines} we show the
evolution with redshift of the CIV cosmological mass density as a
fraction of the critical density today, calculated considering the
column density distribution function of the CIV absorption lines:
\begin{equation}
  \Omega_{\rm CIV}(z)=\frac{H_{\rm 0}\;m_{\rm CIV}}{c\;\rho_{\rm crit}}\int Nf(N)\ dN,
\end{equation}
where $H_{\rm 0}=100$ $h$ km s$^{-1}$ Mpc$^{-1}$ is the Hubble
constant ($h=0.73$), $m_{\rm CIV}$ is the mass of a CIV ion, $c$ is
the speed of light, $\rho_{\rm crit}=1.88\times10^{-29}$ $h^2$ g
cm$^{-3}$ and $f(N)$ is the CIV CDDF. Since $f(N)$ cannot be recovered
correctly for all the column densities due to incompleteness and poor
statistics, the integral in the previous equation can be approximated
by a sum \citep{storrie-lom96}:
\begin{equation}
\label{eq:omciv_sum}
  \Omega_{\rm CIV}(z)=\frac{H_{\rm 0}\;m_{\rm CIV}}{c\;\rho_{\rm crit}}\frac{\sum_{\rm i}N_{\rm i}(CIV)}{\Delta X},
\end{equation}
with $\Delta X\equiv\int(1+z)^2\left[\Omega_{\rm
    0m}(1+z)^3+\Omega_{\rm 0\Lambda}\right]^{-1/2}\ dz$, the redshift
absorption path. In the right panel of Figure \ref{fig:omcivlines} the
overplotted black diamonds show the observational data of
\citet{vale10}. The value of $\Omega_{\rm CIV}$ significantly depends
on the column density range over which the sum or the integration is
carried out and, as a consequence, on the resolution and
signal-to-noise ratio of the available spectra. In order to address
this aspect, \citet{vale10} have computed three sets of values to be
compared consistently with different data in the literature. Here we
compared $\Omega_{\rm CIV}$ obtained from the simulated CIV {\it
  systems} (as defined in Section \ref{obdatasamp}) lying in the
column density range $12\leq\log N_{\rm CIV}\,$(cm$^{-2}$)$\;\leq 15$,
with the $\Omega_{\rm CIV}$ computed from the CIV systems of the total
\citet{vale10} sample, in the same range of column densities. From
$z=3.5$ to $z=2.5$ all the simulations, except for the kr37nf
(no-feedback) and the kr37agn (AGN-feedback), roughly reproduce the
observational data and show an increasing or constant trend for the
$\Omega_{\rm CIV}$. Down to redshift $z=3$ the AGN feedback of run
kr37agn is not active and the CIV content is nearly equal to the one
of the no-feedback simulation kr37nf. At $z=3$ the AGN feedback starts
to work and the kr37agn run suddenly reaches all the other ``wind''
simulations.

Apart from the no-feedback simulation kr37nf (green crosses and dashed
line), all the other runs still reproduce the observational data
around redshift $z=2.25\pm0.25$. If we consider lower redshift, all
the simulations show a decreasing trend at variance with the
data. Different is the case of the kr37agn run: from redshift $z=3.0$
to $1.8$, the AGN feedback suppresses efficiently the star formation,
therefore the gas is no longer reprocessed from the stars and the CIV
is not converted in other ions. At the same time, the AGN feedback has
not yet made the haloes devoid of gas, with the result that $\Omega_{\rm
  CIV}$ continues to increase. However, since the black holes
accretion feedback is extremely powerful, moving to lower redshift a
considerable amount of gas is expelled from the haloes and heated. As
a consequence $\Omega_{\rm CIV}$ starts to decrease, even if at a
later time and to a smaller degree than the ``wind'' runs.

\section{Probability distribution function of the CIV Doppler parameter}
\label{sec:bcivpdf}

\begin{figure*}
\centering
\includegraphics[width=6.5cm]{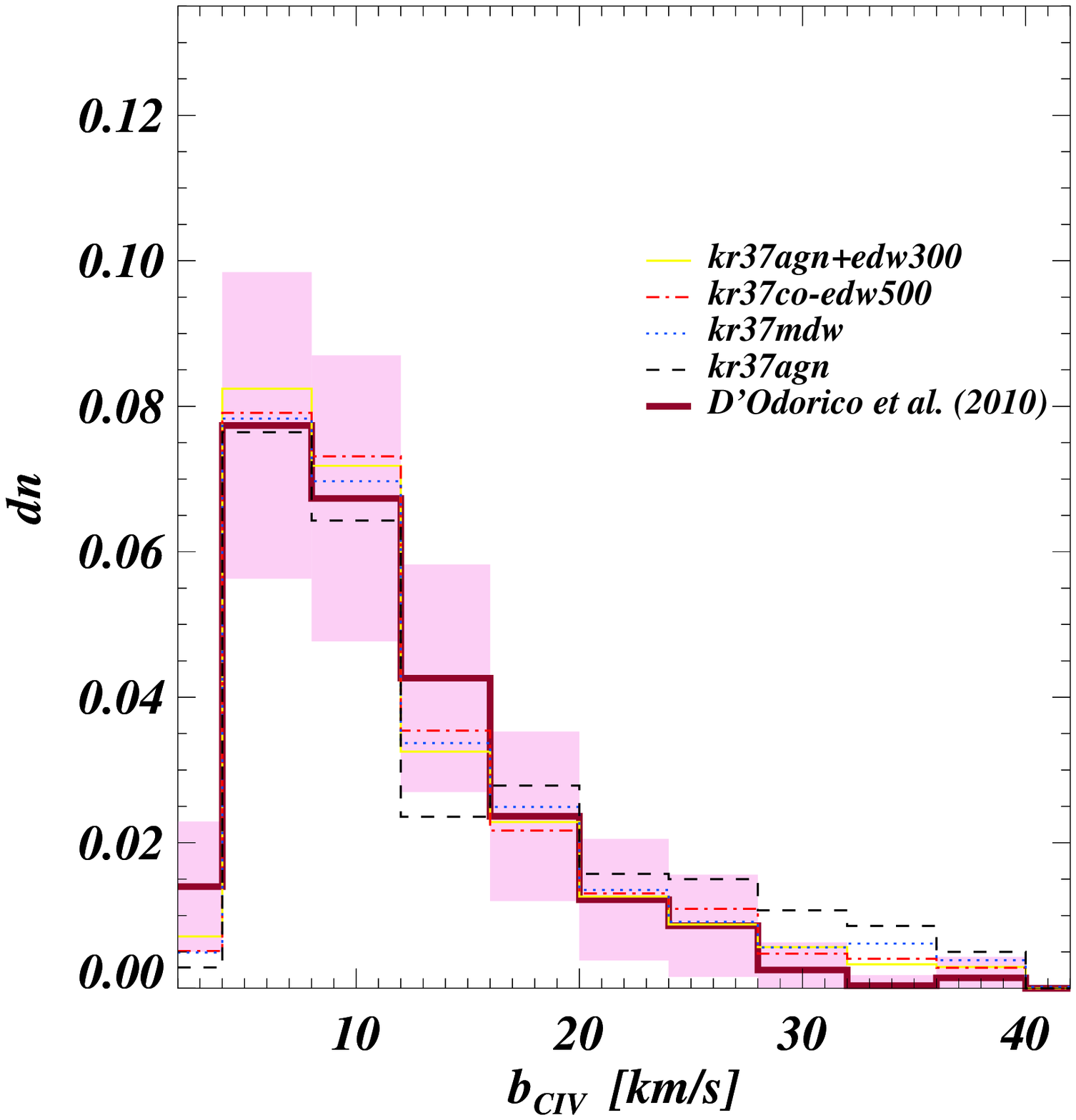}
\includegraphics[width=6.5cm]{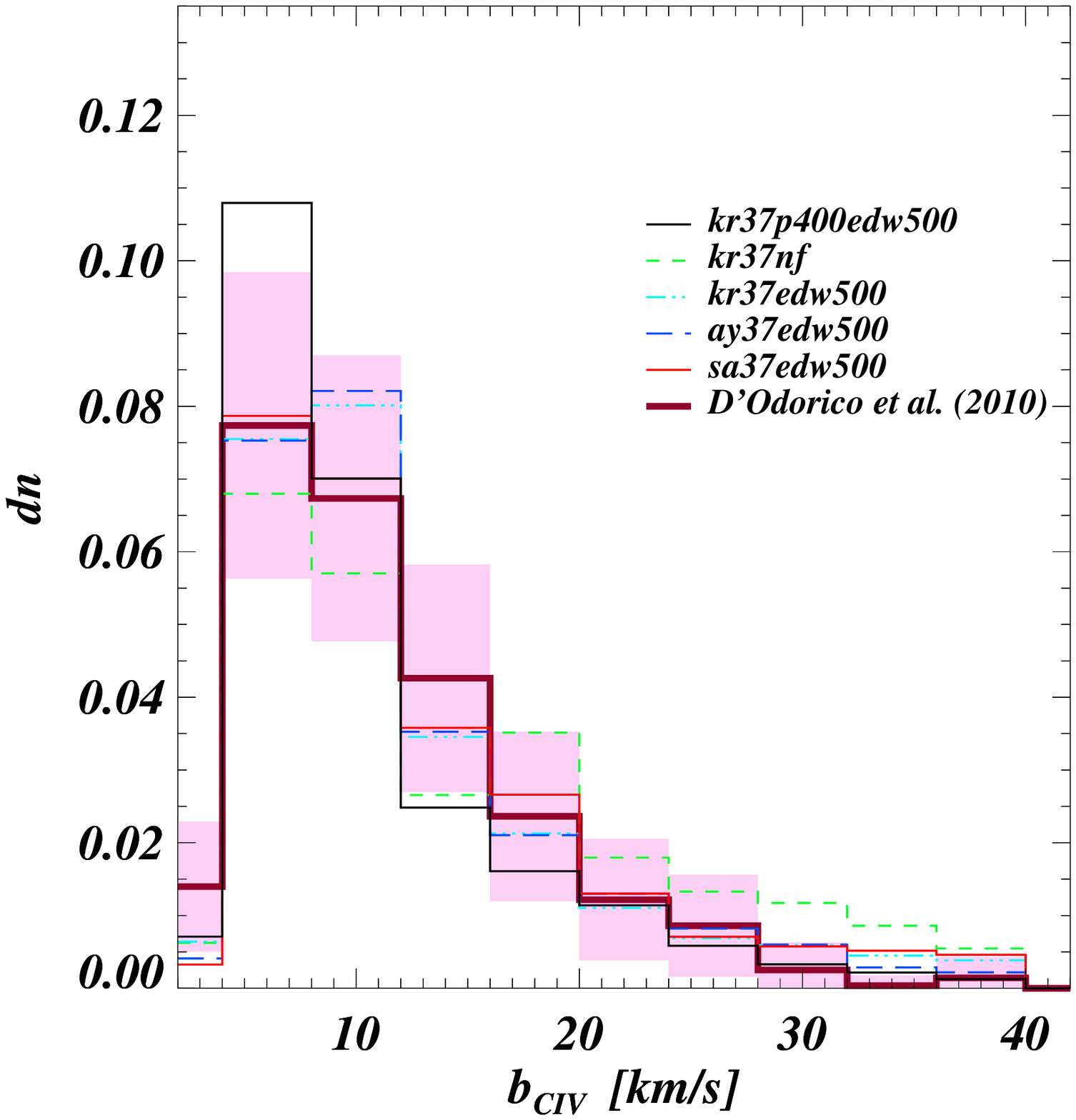}
\caption{Line-widths $b_{\rm CIV}$ probability distribution function
  at $z=3$. \textit{Left panel}: part I. \textit{Right panel}: part
  II. In both panels, data from \citet{vale10} are showed by the
  purple solid line along with the associated Poissonian error (shaded
  region).}
\label{fig:bpdfz3}
\end{figure*}

\begin{figure*}
\centering
\includegraphics[width=6.5cm]{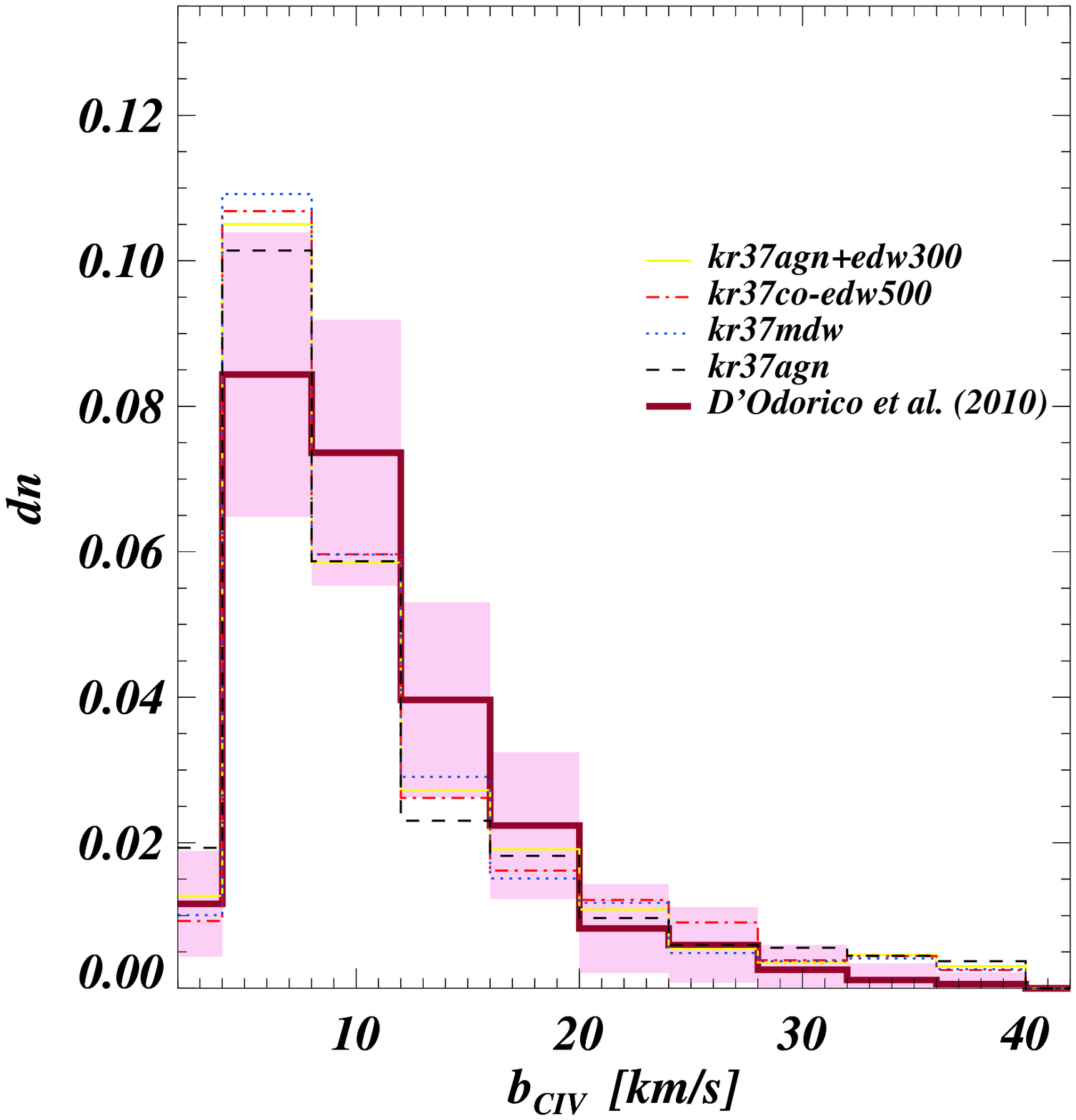}
\includegraphics[width=6.5cm]{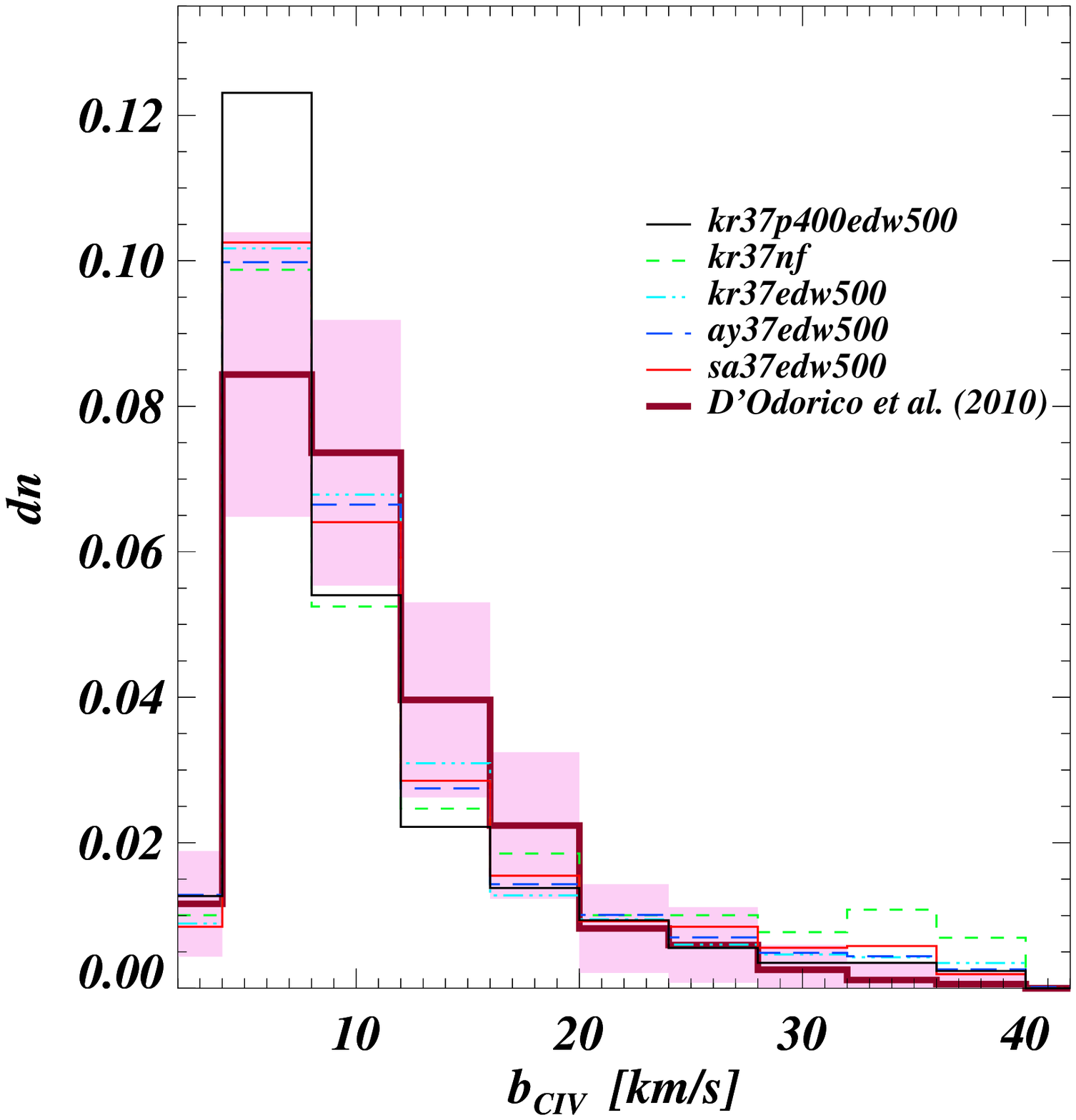}
\caption{As in Figure \ref{fig:bpdfz3}, but at redshift $z=2.25$.}
\label{fig:bpdfz2p25}
\end{figure*}

\begin{figure*}
\centering
\includegraphics[width=6.5cm]{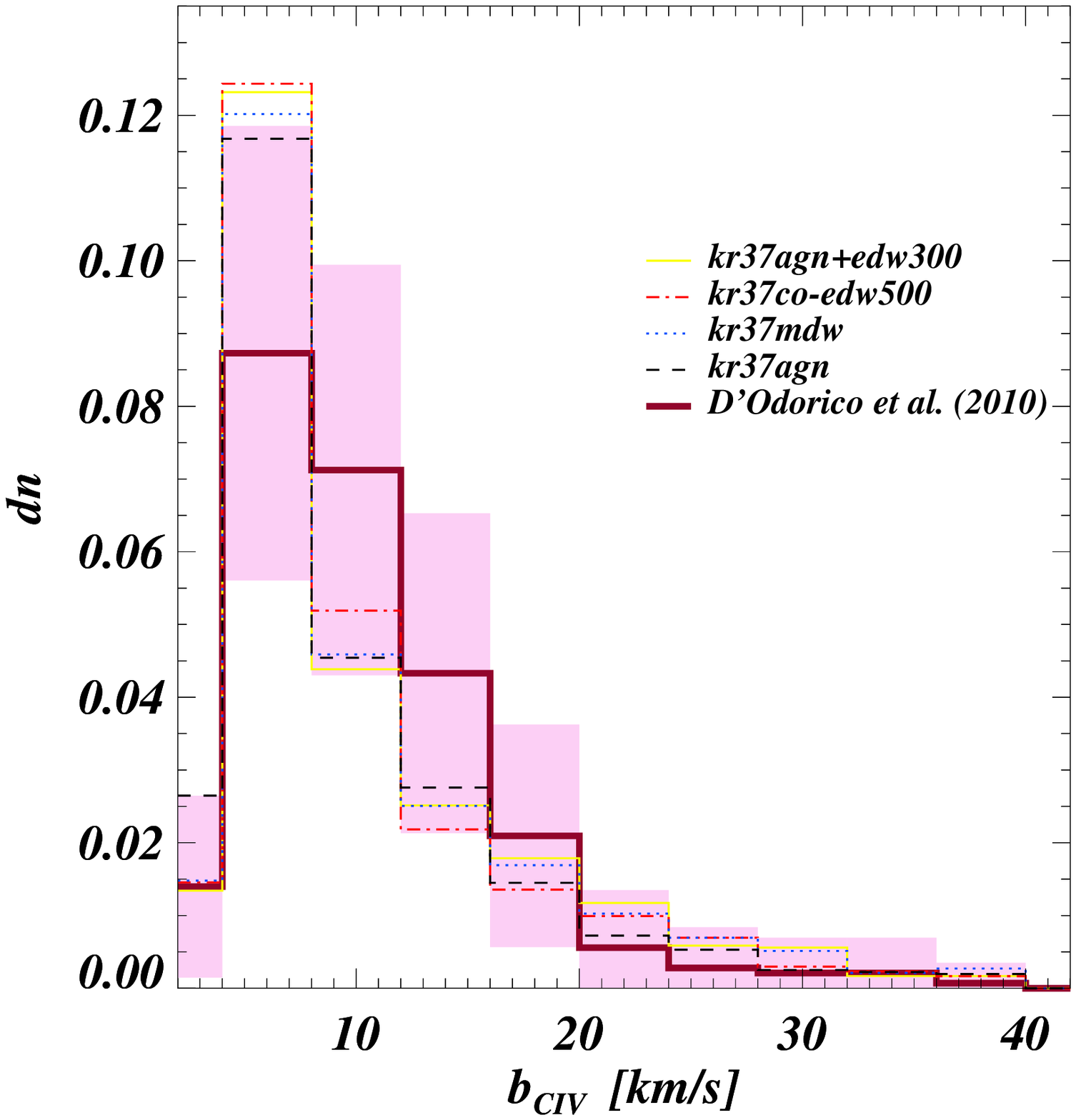}
\includegraphics[width=6.5cm]{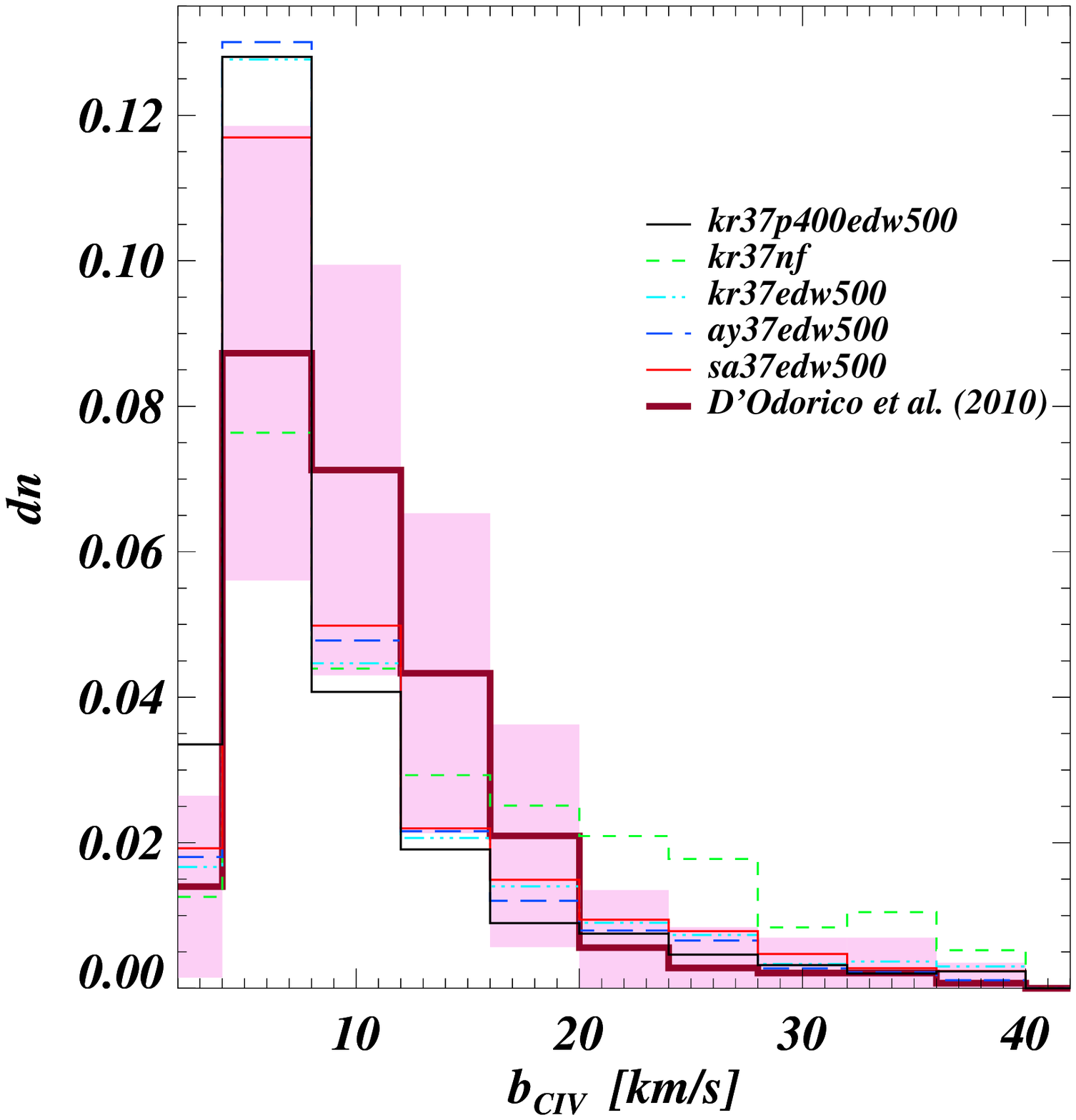}
\caption{As in Figure \ref{fig:bpdfz3}, but at redshift $z=1.5$.}
\label{fig:bpdfz1p5}
\end{figure*}

In this Section we focus on the CIV Doppler parameter probability
distribution function, plotted at redshift $z=3$, 2.25 and 1.5 in
Figures \ref{fig:bpdfz3}, \ref{fig:bpdfz2p25} and \ref{fig:bpdfz1p5},
respectively. The data from \citet{vale10} are also overplotted in
these figures with a purple solid line along with the associated
Poissonian error (shaded region). For this analysis we considered CIV
{\it lines}.

At redshift $z=3$ (Figure \ref{fig:bpdfz3}) all the runs are quite in
agreement with the data even if there are some small
discrepancies. All the runs tend to underestimate the observed
distribution in the first bin, $b_{\rm CIV}<4$ km s$^{-1}$, and
slightly overestimate it for $b_{\rm CIV}>28$ km s$^{-1}$. In the
intermediate range $4<b_{\rm CIV}<28$ km s$^{-1}$, the simulations
behave differently. The kr37mdw, kr37agn+edw300, kr37co-edw500 (left
panel) and sa37edw500 (right panel) runs fit very well the
observational data, while runs kr37edw500 and ay37edw500 (right panel)
overproduce the observed distribution in the bin centered at $b_{\rm
  CIV}=10$ km s$^{-1}$. Again the runs kr37nf and kr37agn show a
different trend with respect to the others, for exactly the same
reasons mentioned in the previous sections. Simulations kr37nf
(no-feedback, right panel green dashed line) and kr37agn
(AGN-feedback, left panel black dashed line) distributions are shifted
towards higher $b_{\rm CIV}$ than the observed one. At this redshift
AGN feedback is not active so, as the no-feedback case of kr37nf, the
gas remains trapped inside haloes at high density and temperature. Run
kr37p400edw500 show a marked excess of low CIV Doppler parameters due
to its improved resolution that can resolve higher densities at
earlier times and produces a lot of small metal enriched clumps in the
IGM. At $z=3$, the median value of the observational Doppler
parameters distribution is 9.88 km s$^{-1}$, while the simulations
have median ranging from 8.49 km s$^{-1}$ (kr37p400edw500) to 13.03 km
s$^{-1}$ (kr37nf).

\begin{figure*}
\centering
\includegraphics[scale=0.54]{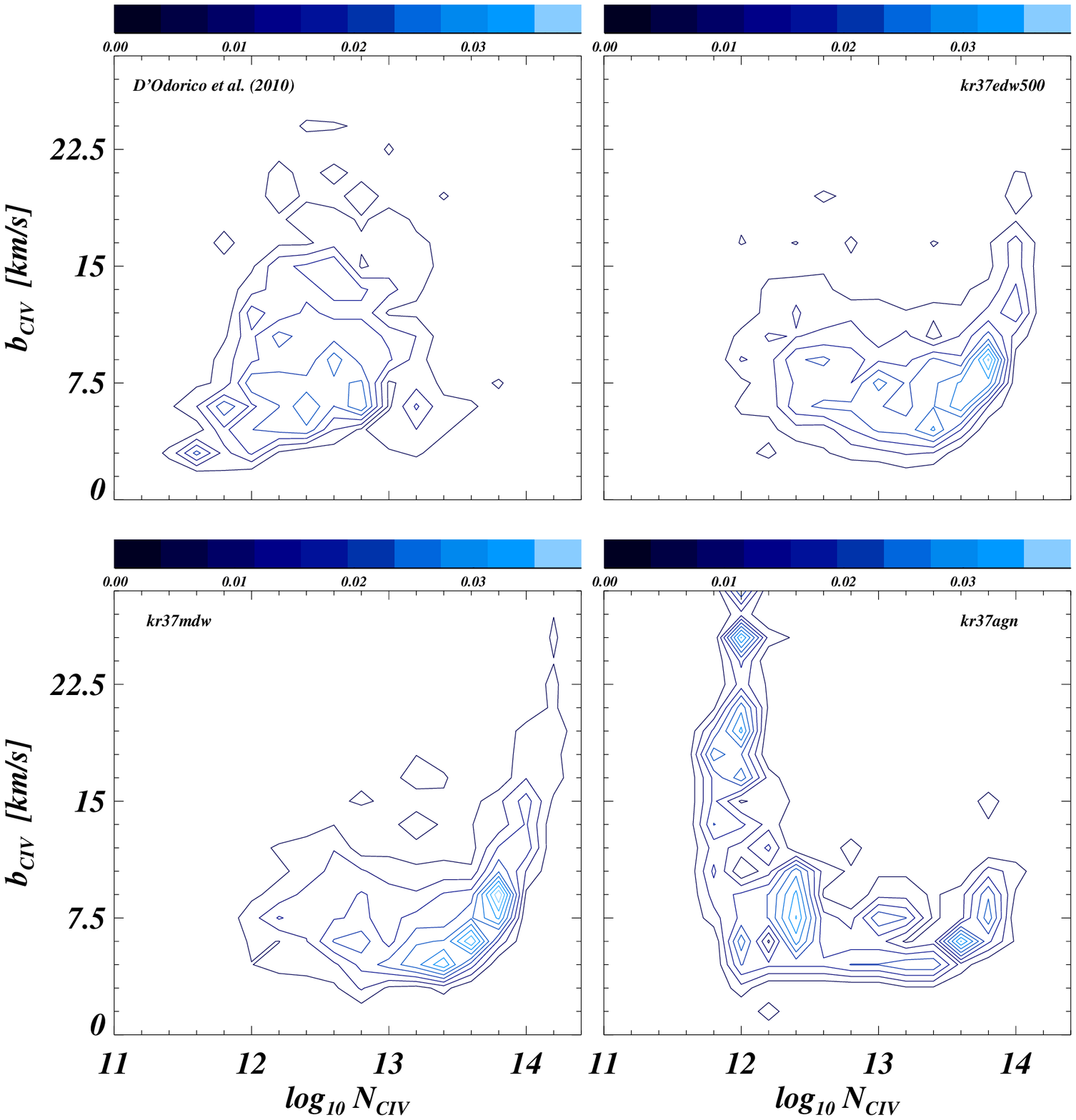}
\caption{$b_{\rm CIV}-N_{\rm CIV}$ relation at $z=3$. The horizontal
  bars are colour coded according to the fraction of points that fall
  in each bin with coordinate ($b_{\rm CIV}$,$N_{\rm CIV}$). Upper
  left panel: observational data from \citet{vale10}.}
\label{fig:bNc4_z3}
\end{figure*}

The situation is different at redshift $z=2.25$ (Figure
\ref{fig:bpdfz2p25}) and $z=1.5$ (Figure \ref{fig:bpdfz1p5}). Moving
to lower redshift, all the simulations show an excess of low Doppler
parameters ($b_{\rm CIV}<8$ km s$^{-1}$), and also they underproduce
the observed distribution in the range $8<b_{\rm CIV}<20$ km
s$^{-1}$. Moreover, while at low redshift the kr37nf simulation
(no-feedback, right panels green dashed lines) is still shifted
towards higher $b_{\rm CIV}$ with respect to the observed
distribution, the kr37agn run (AGN-feedback, left panels black dashed
lines) at redshift $z=2.25$ fits better the data and it finally
approaches all the other runs at $z=1.5$, further confirming the
increasing efficiency of the AGN feedback at low redshift. At $z=2.25$
and $z=1.5$, the data by \citet{vale10} have median value respectively
equal to 9.30 and 9.25 km s$^{-1}$, while the simulations have median
around, respectively, 8.55 and 7.40 km s$^{-1}$ (12.47 km s$^{-1}$ in
the case of the no-feedback run kr37nf).

We stress that the small discrepancy between observed and simulated
distributions for low Doppler parameters ($b_{\rm CIV}<8$ km s$^{-1}$)
appears in a range that is influenced by the adopted resolution
(remember that we have convolved the simulated spectra with a Gaussian
of 6.6 km s$^{-1}$ FWHM) and a sightly different value than the one
used could alleviate this discrepancy

Compared to that of HI, the CIV Doppler parameter distribution
function is less affected by the choice of the UV background (and in
particular of the factor of 3.3 discussed in Sections \ref{sec_sim} and
\ref{sec:bHI_pdf}). This is due to the fact that CIV absorption is
related to higher density regions (mainly the outskirts of galactic
haloes) than the HI absorption. In these regions, the impact of the
IGM local temperature and density on the ionization state of Carbon is
considerably more important than the UVB contribution.

\section{The CIV column density-Doppler parameter relation}
\label{sec:bcivNciv}

\begin{figure*}
\centering
\includegraphics[scale=0.54]{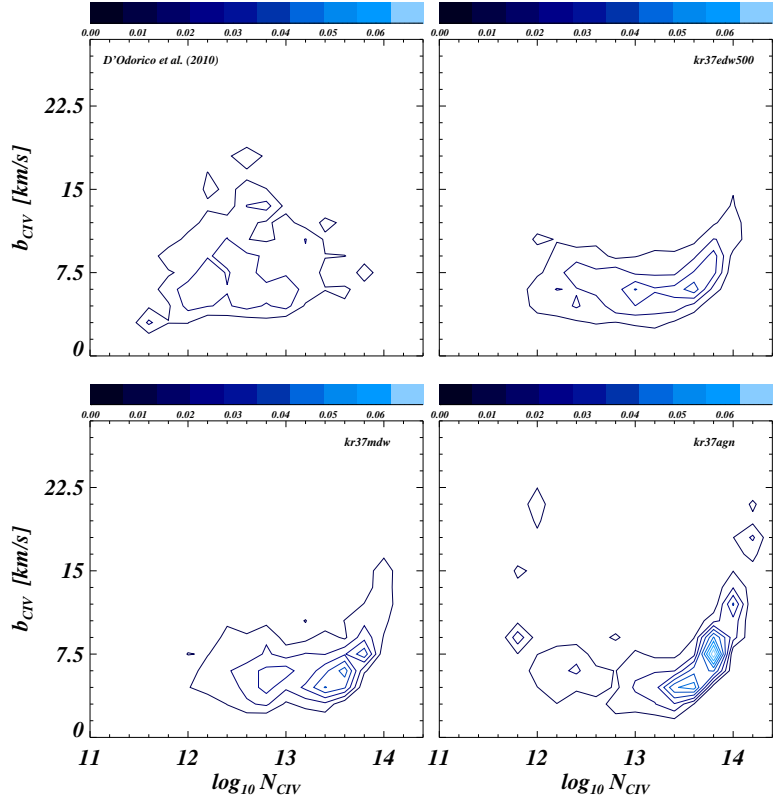}
\caption{As in Figure \ref{fig:bNc4_z3}, but at redshift $z=2.25$.}
\label{fig:bNc4_z2p25}
\end{figure*}

\begin{figure*}
\centering
\includegraphics[scale=0.54]{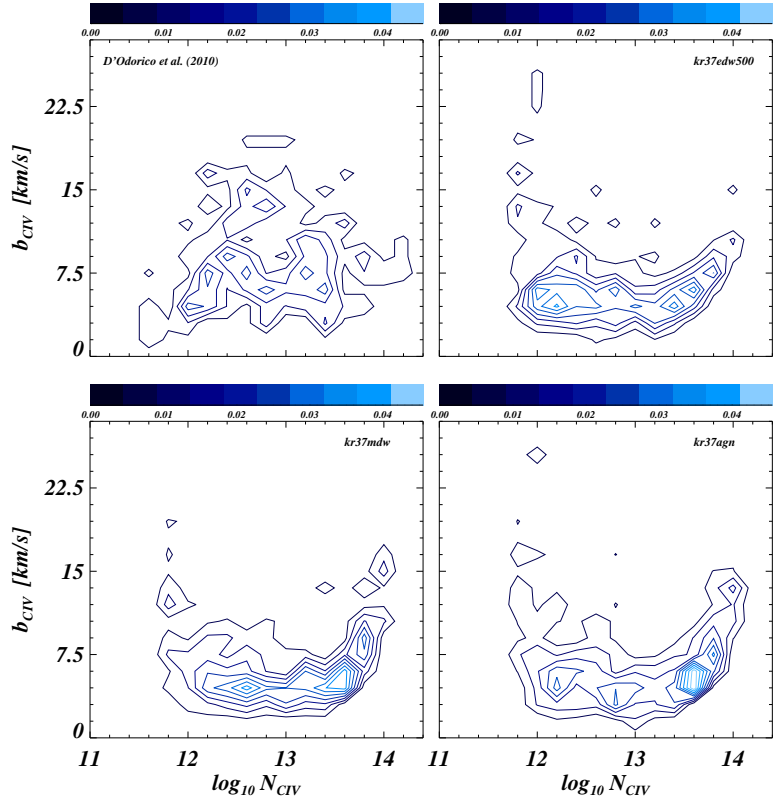}
\caption{As in Figure \ref{fig:bNc4_z3}, but at redshift $z=1.5$.}
\label{fig:bNc4_z1p5}
\end{figure*}

In Figures \ref{fig:bNc4_z3}, \ref{fig:bNc4_z2p25} and
\ref{fig:bNc4_z1p5} we plot the $b_{\rm CIV}-N_{\rm CIV}$ relation, at
redshift $z=3$, $z=2.25$ and $z=1.5$ respectively, for the reference
simulations kr37edw500, kr37mdw and kr37agn. The contour plots are
colour coded according to the fraction of points that fall in each bin
with coordinate ($b_{\rm CIV}$,$N_{\rm CIV}$). In the upper left panel
of the figures the observational data of \citet{vale10} are
plotted. As in the previous Section, we considered here CIV {\it
  lines}.

At redshift $z=3$ (Figure \ref{fig:bNc4_z3}), the data distributions
of runs kr37edw500 (energy-driven winds) and kr37mdw (momentum-driven
winds) are in agreement with each other but are quite different
from the observed one. In particular the simulated distributions are
shifted towards higher CIV column densities and are concentrated
around a narrower region than the observational data. The low column
density tail of the distributions is completely missing, while the
high column density tail shows a sort of correlation between the CIV
Doppler parameters and $N_{\rm CIV}$. We interpret this latter as a
$T-\rho$ relation: lines with higher column densities (i.e. associated
to denser regions) have higher Doppler parameters
(i.e. temperature). The fact that this relation appears only at high
column densities is not a surprise: for high values of $N_{\rm CIV}$,
absorption lines are very strong and well resolved, so they are good
tracers of the physical state of the intergalactic gas and also they
are better fit by {\small VPFIT}. The case of run kr37agn is
different: in fact the $T-\rho$ relation for large $N_{\rm CIV}$
values is missing, while a clump of systems with a large spread in
$b_{\rm CIV}$ values at low column density is clearly visible. This
distribution results because, at redshift $z=3$, in the kr37agn run
the gas resides inside the high density high temperature cores of the
haloes where it is reprocessed by the stars, and the amount of diffused
CIV is rather low (see the right panel of Figure
\ref{fig:omcivlines}). Therefore, since the CIV absorption lines are
weak and not well defined, the quality of the fit made by {\small
  VPFIT} is rather poor. As we reported in Section \ref{sec:bHI_pdf},
in such a case {\small VPFIT} tends to add broad components in order
to minimize the $\chi^{2}$ statistics and this results in the feature
of spurious systems at low column density shown in the lower right
panel of Figure \ref{fig:bNc4_z3}. While the effect is less evident
(but present) for the other runs or in the case of the $b_{\rm
  HI}-N_{\rm HI}$ relation, it is now much more prominent. Of course,
observational data do not present such a feature, because in the ``by
eye'' procedure of fitting these spurious lines are removed.

At redshift $z=2.25$ (Figure \ref{fig:bNc4_z2p25}) and $z=1.5$ (Figure
\ref{fig:bNc4_z1p5}) the kr37agn run approaches the other two
simulations. The $b_{\rm CIV}-N_{\rm CIV}$ correlation at high column
densities is now present for all the runs but the simulated
distributions are still different from the observed one. The clump of
low density systems disappears almost completely at $z=2.25$, but is
visible also for the runs kr37edw500 and kr37mdw at redshift
$z=1.5$. There is a simple reason for this: at low redshift the amount
of CIV, traced by random line-of-sight along the box, decreases (right
panel of Figure \ref{fig:omcivlines}), while the number of low column
density systems significantly increases (Figure \ref{fig:cddfz1p5})
and typically these lines have a large spread in their Doppler
parameter value.

The difference between simulated and observed distributions could also
be due to the fact that simulated haloes, and in particular their
outskirts, can have different properties from that of the observed ones: for
example this is also seen from the SiII statistics in DLA systems that
appear to be not in agreement with observations. Although in the
present paper we explore the effect of feedback, there might be other
physical effects that can contribute as well in shaping the properties
of the $b_{\rm CIV}-N_{\rm CIV}$ distribution.

\begin{figure*}
\centering \includegraphics[scale=0.7]{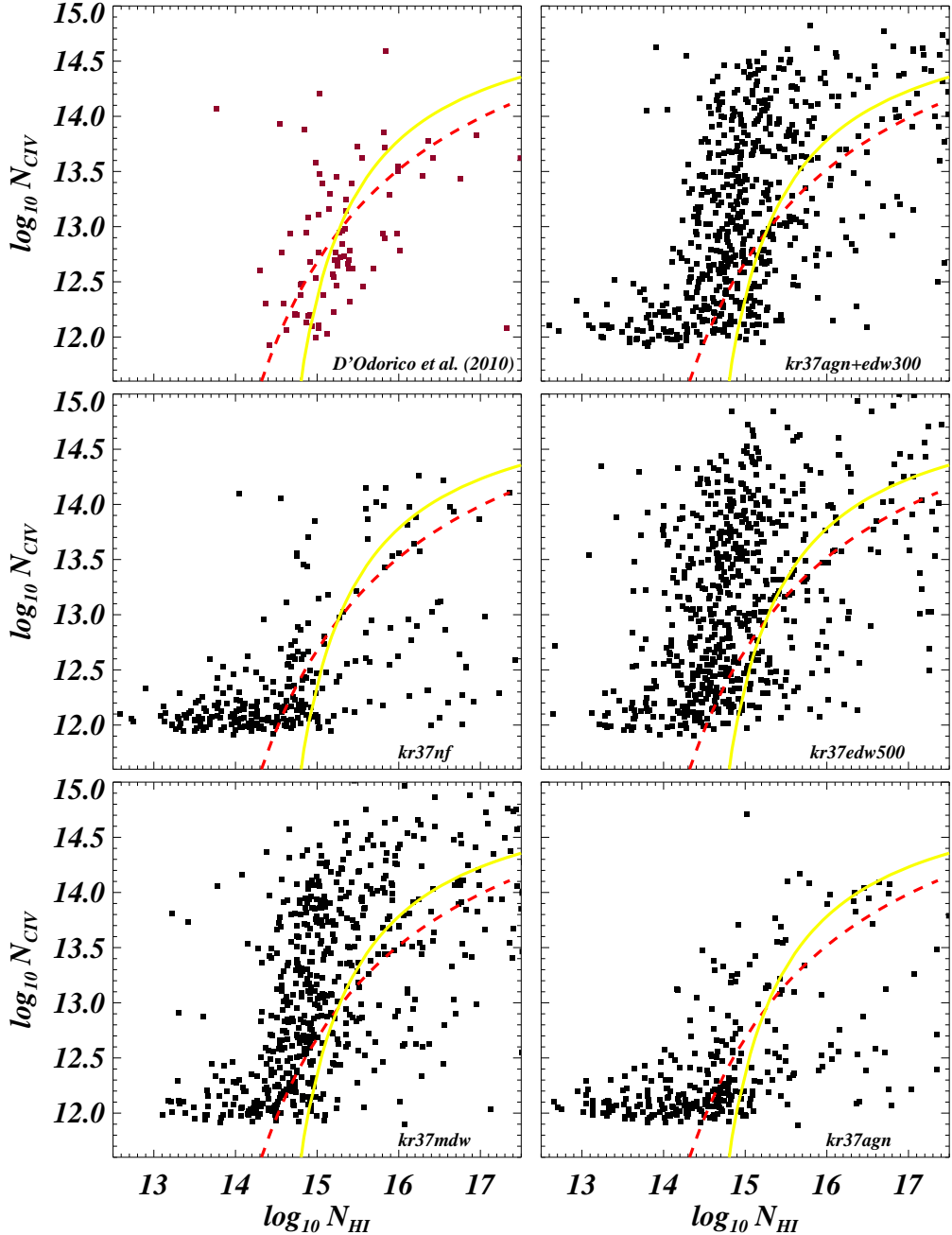}
\caption{Column densities $N_{\rm CIV}-N_{\rm HI}$ correlated
  absorption for systems at $z=3$. Upper left panel: observational
  data from \citet{vale10}. The red-dashed and yellow-solid curves
  represent the fitting functions of \citet{kimsub}.}
\label{fig:systems_z3}
\end{figure*}

\begin{figure*}
\centering
\includegraphics[scale=0.7]{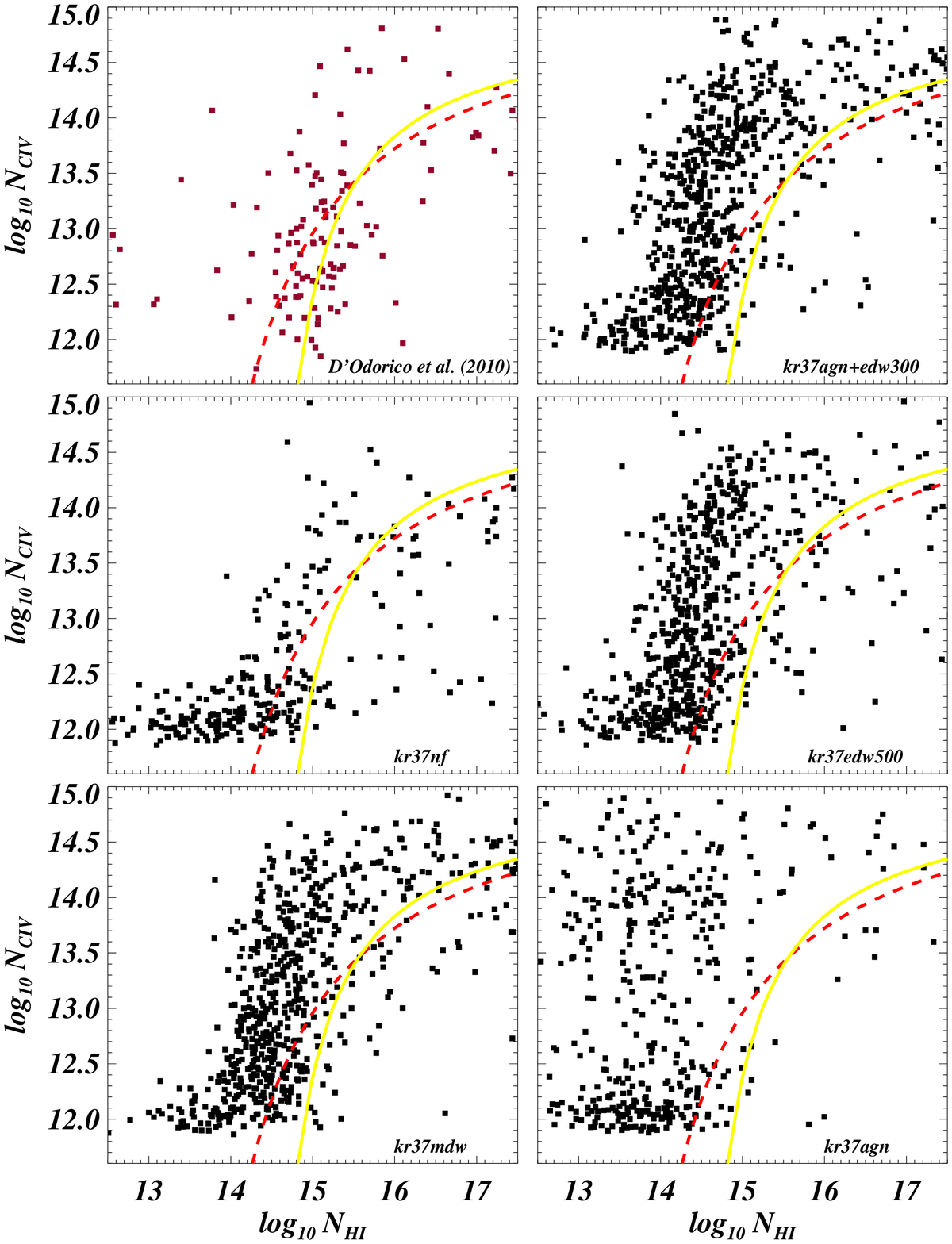}
\caption{As in Figure \ref{fig:systems_z3}, but at redshift $z=2.25$.}
\label{fig:systems_z2p25}
\end{figure*}

\section{$N_{\rm CIV}-N_{\rm HI}$ correlated absorption}
\label{sec_corrCIVHI}

The last part of this work is dedicated to the HI-CIV correlated
absorption. We define as correlated absorption HI-CIV systems in which
CIV and HI are physically dependent or, in other words, related to the
same absorptive structure. In real spectra, however, two absorbers
each containing CIV could be very close in redshift space but far in
real space due to the bulk motions and the peculiar velocity. For this
reason there are often serious problem on how to associate each CIV
components to physically corresponding HI components. In this Section,
besides the data of \citet{vale10}, we compare with the fitting
formulae obtained from the \citet{kimsub} data. In oder to do that,
only for this analysis we changed the velocity range, $dv_{\rm min}$,
used so far to define {\it systems} of lines. Following \citet{kimsub}
and differently from what was done in Sections \ref{sec_civcddf} and
\ref{sec:omcivev}, where we set $dv_{\rm min} = 50$ km s$^{-1}$, we
use here the velocity interval [-250 km s$^{-1}$, 250 km s$^{-1}$]
both for the simulated and the observed CIV and HI systems.

In Figures \ref{fig:systems_z3} and \ref{fig:systems_z2p25} we show
our findings for some of the simulations in Table \ref{tab:sim_civ}
and for the \citet{vale10} data. The red-dashed and yellow-solid
curves represent the \citet{kimsub} fit, obtained by using the
function:
\begin{equation}
  \log N_{\rm CIV}=\left[\frac{C_{\rm 1}}{\log N_{\rm HI}+C_{\rm
        2}}\right]+C_{\rm 3}.
\end{equation}
The parameters for the red-dashed curves are: $C_{\rm
  1}=[-4.43\pm1.30,-7.57\pm1.80]$, $C_{\rm
  2}=[-13.04\pm0.44,-12.45\pm0.45]$, $C_{\rm
  3}=[15.22\pm0.17,15.65\pm0.19]$ at $z=[2.25,3.0]$. These red-dashed
curves represent a fit to {\it all} the \citet{kimsub} data available
at the two different redshifts. The parameters for the yellow-solid
curves are: $C_{\rm 1}=[-2.01\pm0.35,-2.51\pm0.47]$, $C_{\rm
  2}=[-14.22\pm0.14,-14.08\pm0.20]$, $C_{\rm
  3}=[14.96\pm0.05,15.09\pm0.07]$ at $z=[2.25,3.0]$. These curves
represent an improved fit made to reproduce better the clump of
systems at low $N_{\rm HI}$ and low $N_{\rm CIV}$.

At redshift $z=3$ (Figure \ref{fig:systems_z3}) for the \citet{vale10}
observational data and for the kr37agn+edw300, kr37edw500 and kr37mdw
runs the bulk of systems is located at $\log N_{\rm
  HI}\,$(cm$^{-2}$)$\;<16$ and it follows the \citet{kimsub} fitting
functions, even if with a slightly steeper trend, while at $\log
N_{\rm HI} \,$(cm$^{-2}$)$\;>16$ the spread in the simulated data
increases \citep[also the data of][present a sort of bimodality with
  high HI column density distribution more spread than the tight low
  HI column density one]{kimsub}. Runs kr37nf (no-feedback) and
kr37agn (AGN-feedback) at this redshift behave similarly: they do not
show the $\log N_{\rm HI}\,$(cm$^{-2}$)$\;<16$ CIV-HI correlation
above, and the $\log N_{\rm HI}\,$(cm$^{-2}$)$\;>16$ systems are
rare. All the simulations show a strip of values with $\log N_{\rm
  HI}\,$(cm$^{-2}$)$\;<15$ and $\log N_{\rm
  CIV}\,$(cm$^{-2}$)$\;\sim12$. This is due to the {\small VPFIT}
spurious line fitting effect described in the previous Section. In
fact, it is particular evident for the runs kr37nf and kr37agn,
because in these two cases the amount of the diffuse CIV is lower and
the CIV absorption lines are weak and not well defined.

At redshift $z=2.25$ (Figure \ref{fig:systems_z2p25}) the above trends
are in place even if all the simulated distributions are now shifted
towards slightly lower HI column densities with respect to the fitting
functions of \citet{kimsub}. Otherwise, the number of components of
the \citet{vale10} data increases at this redshift and the bulk of
systems at $\log N_{\rm HI}\,$(cm$^{-2}$)$\;<16$ is well fitted by the
functions of \citet{kimsub}. Again, moving to lower redshift, the
simulation with AGN feedback, kr37agn, starts to follow the other wind
runs.

To summarize, two effects are visible from the plots: $(a)$ for a
given $N_{\rm HI}$, the scatter in the corresponding $N_{\rm CIV}$ is
broader, and the $N_{\rm CIV}$ values typically larger, than what is
inferred from observations; $(b)$ the simulated distributions of $N_{\rm HI}$ are
slightly lower compared to the \citet{kimsub} fitting functions and
the \citet{vale10} data at redshift $z=2.25$.

\section{Conclusions and Discussion}
\label{concl}
In this paper we investigated the global properties of triply ionized
Carbon (CIV) as an IGM tracer at $z\geq1.5$. From the numerical point
of view we presented the results of a new set of hydrodynamic
simulations that incorporate feedback either in the form of galactic
winds or in the form of energy from black holes accretion (AGN
feedback). From the observational point of view we relied on recent
high-resolution data sets obtained with the UVES spectrograph at VLT
and the HIRES spectrograph at Keck. The main results can be summarized
as follows:
\begin{itemize}
\item The statistics of HI are weakly affected by the feedback
  prescription implemented: both the column density distribution
  function and the Doppler width distributions do not change
  significantly when considering the different simulations. We regard
  this as a geometrical effect: winds and AGN feedback are stronger at
  the intersection of IGM filaments and these sites have usually small
  filling factor. Only the high column density tail shows small
  differences between the runs considered (Section
  \ref{sec_histati}). The fact that overall our simulations reproduce
  well all the HI statistics confirms that we are catching the physics
  of the gas traced by the neutral hydrogen. The remaining small
  discrepancies with the observational data could be due to the UV
  background we used \citep[][produced by quasars and galaxies, with
    the heating rates multiplied by a factor of $3.3$]{HM96}, that heats
  too much the gas. In minor part, the discrepancies are also due to a
  numerical effect introduced by the line-fitting software {\small
    VPFIT}.
\item Compared to that of HI, the statistics of CIV is strongly
  affected by feedback. If we consider {\it systems} of lines as
  defined in Section \ref{obdatasamp}, at redshift $z=3$, 2.25 and
  1.5, all the simulations except for the no-feedback run kr37nf and
  the AGN-feedback run kr37agn (this latter only at high redshift)
  agree with the observed CIV column density distribution function. Galactic
  winds feedback starts to be active at high redshift, but moving to
  lower redshift, also the AGN feedback becomes effective and the
  kr37agn simulation follows all the other wind runs (see Figures
  \ref{fig:cddfz3}, \ref{fig:cddfz2p25} and \ref{fig:cddfz1p5}).
\item In the second part of Section \ref{sec:omcivev}, we presented
  the evolution with redshift of the CIV cosmological mass density,
  calculated considering the column density distribution function of
  the CIV absorption lines. In the redshift range $z=2.5-3.5$, our
  simulations, except for the no-feedback run kr37nf and the
  AGN-feedback run kr37agn, reproduce the observed $\Omega_{\rm
    CIV}(z)$ evolution, even if with a slightly overestimation, and
  they perfectly reproduce the observational data around redshift
  $z=2.25\pm0.25$. At lower redshift we found a decreasing trend at
  variance with the increasing trend shown by the observational data
  (right panel of Figure \ref{fig:omcivlines}). In the AGN case the
  trend is nearly constant at low redshift and in better agreement
  with data. We think that a better calibration of the coupled AGN +
  galactic winds feedback model of run kr37agn+edw300 could help to
  improve this result.
\item At all the redshift considered, the CIV Doppler parameter
  distribution is in good agreement with the observational data. The
  no-feedback and AGN-feedback simulations (the latter only at high
  redshift) result in distributions shifted towards higher $b_{\rm
    CIV}$ than the observed ones: this is due to the fact that these
  runs produce gas around galaxies that is too hot than what was inferred
  from observations (Section \ref{sec:bcivpdf}).
\item The overall $b_{\rm CIV}-N_{\rm CIV}$ data distribution of our
  reference simulations is different from the observed one. The high
  column density tail of all the runs considered (but for the
  AGN-feedback run this is true only at low redshift) shows a sort of
  correlation between the CIV Doppler parameters and the column
  density, in which the Doppler parameters increase as $N_{\rm CIV}$
  increases: the sign of a well defined temperature-density relation
  for the high density gas. Especially at low redshift, when the
  amount of CIV traced by random line-of-sight along the box
  decreases, a clump of systems with a huge scatter in the $b_{\rm
    CIV}$ values at low CIV column density is present. This effect
  should be regarded as of numerical origin and it is introduced by
  {\small VPFIT} when fitting regions close to the continuum in the
  mock data set (Section \ref{sec:bcivNciv}). A more careful
  comparison between data and simulations is needed in order to
  address this effect at a more quantitative level.
\item In Section \ref{sec_corrCIVHI} we explored the correlated CIV-HI
  absorption, considering {\it systems} of lines in which CIV and HI
  are physically dependent. At the two redshift considered, $z=3$ and
  $z=2.25$, both our reference runs and the observational data of
  \cite{vale10} roughly follow the fit proposed by \citet{kimsub},
  even if the simulations show a slightly steeper trend in the range
  $\log N_{\rm HI} \,$(cm$^{-2}$)$\;< 16$.
\item Even if we are not able to fit all the statistics at once if we
  consider the neutral hydrogen HI and the triply ionized Carbon CIV,
  momentum-driven winds simulation reproduce best all the different
  quantities explored in this paper (as found for Damped
  Lyman-$\alpha$ systems by Tescari et al. 2009, and for IGM metal
  lines by Oppenheimer \& Dav\'e 2006, 2008).
\item To sum up, feedback appears to be a crucial physical ingredient
  in order to reproduce statistics of metal absorption lines. In
  particular, the effect of galactic winds triggered by supernovae is
  important at high redshift, while the effect of black holes
  accretion feedback starts to become prominent at $z<3$. However,
  because the AGN feedback is very strong, it will be crucial to
  improve the coupled AGN + galactic winds model, where the effect of
  the winds at high redshift reduces the strength of the black holes
  accretion at low redshift.
\end{itemize}

In our analysis, we found two different effects biasing the statistics
related to the HI and CIV absorption lines: the choice of the UV
background and the {\small VPFIT} lines fitting procedure. In the
future we will try to improve these statistics by using a more
physically motivated UV background that takes into account HeII
reionization at $z\sim 3$. Furthermore, we will try to improve our
automatic line-fitting analysis by removing numerical artifacts
introduced by {\small VPFIT}, since the automatic fitting procedure is
not likely to be as accurate as a ``by eye'' fitting made by an
observer, and introduce spurious lines especially in the continuum
region. Our results can also be tested by using other statistics based
on pixel-optical depths techniques.

A second possible improvement of the results presented in this paper
is the refinement of the feedback models and the inclusion of new
physical processes like turbulence, metal diffusion and radiative
transfer, that can help in polluting with metals the low-density
IGM. So far we explored feedback prescriptions in the form of
energy-driven galactic winds (EDW), momentum-driven galactic winds
(MDW) and black holes accretion (AGN feedback). Exploring the combined
effect of these models seems to be promising: for example the ``energy
coupled momentum driven winds'' model by \citet{choi2010} succeeds in
enriching the gas (like the EDW model) without heating too much the
IGM (like the MDW model). The modelization inside the hydrodynamic
simulations of the radiative transfer is fundamental and could also
impact on the metal mixing. Finally, it will be crucial to incorporate
in the {\small GADGET-2} code the small-scale turbulence and its
impact on the metal diffusion at large scales. With physical mixing,
fluid elements on a fixed (resolved) physical scale do exchange
energy/entropy due to unresolved (turbulent) motions: diffusion allows
some ejecta gas to mix while exiting the galaxies
\citep{shenetal09}. All the previous effects, once included in
simulations, could help in improving our understanding of the chemical
and physical evolution of the IGM and provide a more comprehensive
framework of the high-redshift galaxy/IGM interplay.

\section*{Acknowledgments}
Numerical computations were done on the COSMOS (SGI Altix 3700)
supercomputer at DAMTP and at High Performance Computer Cluster (HPCF)
in Cambridge (UK) and at CINECA (Italy).  COSMOS is a UK-CCC facility
which is supported by HEFCE, PPARC and Silicon Graphics/Cray Research.
The CINECA (``Centro Interuniversitario del Nord Est per il Calcolo
Elettronico'') CPU time has been assigned thanks to an INAF-CINECA
grant. This work has been partially supported by the INFN-PD51 grant,
an ASI-AAE Theory grant and a PRIN-MIUR.

\bibliographystyle{mn2e}
\bibliography{tescari_civ_proof}

\label{lastpage}
\end{document}